\documentclass[twocolumn,tighten,times]{aastex63}
\usepackage{graphicx}
\usepackage{amsmath}
\usepackage{natbib}
\usepackage{amssymb}

\usepackage{sidecap}

\shorttitle{QPEs from Interacting EMRIs }
\shortauthors{Metzger, Stone \& Gilbaum}

\begin{document}

\newcommand{\be}{\begin{equation}}
\newcommand{\ee}{\end{equation}}

\title{Interacting Stellar EMRIs as Sources of Quasi-Periodic Eruptions in Galactic Nuclei}

\author[0000-0002-4670-7509]{Brian D. Metzger}
\affil{Department of Physics and Columbia Astrophysics Laboratory, Columbia University, Pupin Hall, New York, NY 10027, USA}
\affil{Center for Computational Astrophysics, Flatiron Institute, 162 5th Ave, New York, NY 10010, USA} 

\author[0000-0002-4337-9458]{Nicholas C.~Stone}
\affil{Racah Institute of Physics, The Hebrew University, Jerusalem, 91904, Israel}

\author{Shmuel Gilbaum}
\affil{Racah Institute of Physics, The Hebrew University, Jerusalem, 91904, Israel}

\begin{abstract}
A star that approaches a supermassive black hole (SMBH) on a circular extreme mass ratio inspiral (EMRI) can undergo Roche lobe overflow (RLOF), resulting in a phase of long-lived mass-transfer onto the SMBH.  If the interval separating consecutive EMRIs is less than the mass-transfer timescale driven by gravitational wave emission (typically $\sim 1-10$ Myr), the semi-major axes of the two stars will approach each another on scales of $\lesssim$ hundreds to thousands of gravitational radii.  Close flybys tidally strip gas from one or both RLOFing stars, briefly enhancing the mass-transfer rate onto the SMBH and giving rise to a flare of transient X-ray emission.  If both stars reside in an common orbital plane, these close interactions will repeat on a timescale as short as hours, generating a periodic series of flares with properties (amplitudes, timescales, sources lifetimes) remarkably similar to the ``quasi-periodic eruptions'' (QPEs) recently observed from galactic nuclei hosting low-mass SMBHs.  A cessation of QPE activity is predicted on a timescale of months to years, due to nodal precession of the EMRI orbits out of alignment by the SMBH spin.  Channels for generating the requisite coplanar EMRIs include the tidal separation of binaries (Hills mechanism) or Type I inwards migration through a gaseous AGN disk.  Alternative scenarios for QPEs, that invoke single stellar EMRIs on an eccentric orbit undergoing a runaway sequence of RLOF events, are strongly disfavored by formation rate constraints.  
\end{abstract}


\section{Introduction}

Quasi-periodic eruptions (QPEs) are a newly discovered class of short X-ray bursts that originate in spatial coincidence with galactic nuclei, both active and otherwise inactive.  They last for a duration $\tau_{\rm QPE} \lesssim$ hrs, recur with periods, $T_{\rm QPE}$, that range from hours to almost a day between different sources, and exhibit peak luminosities at least an order of magnitude above the quiescent level (\citealt{Miniutti+19,Giustini+20,Arcodia+21}).   

The first QPE system, GSN 069, discovered with {\it XMM-Newton}, exhibited bursts with a recurrence time period $T_{\rm QPE} \approx 8.3$ hr, which increased to $\approx 9.2$ hr over observations spanning several months.  A second QPE system was discovered in RX J1301.9+2747 \citep{Giustini+20}, for which $T_{\rm QPE} \approx 3.6-5.6$ hr.  Two more QPE systems, eRO-QPE1 and eRO-QPE2, were recently discovered with the {\it eROSITA} instrument \citep{Predehl+21} on the Spectrum-Roentgen-Gamma (SRG; \citealt{Sunyaev+21}) space observatory \citep{Arcodia+21}.  These sources were later monitored in follow-up observations by {\it XMM-Newton} and {\it NICER}, the latter detecting 15 consecutive eruptions over 11 days in eRO-QPE1.  The eruptions from eRO-QPE1 (eRO-QPE2) exhibit mean durations of $\tau_{\rm QPE} \approx 7.6 (0.44)$ hr and recurrence periods $T_{\rm QPE} \approx 18.5 (2.4)$ hr \citep{Arcodia+21}, bracketing the range observed in GSN 069 and RX J1301.9+2747.

QPEs exhibit peak X-ray luminosities $L_X \sim 10^{42}-10^{43}$ erg s$^{-1}$ in the $\approx 0.5-2$ keV band with a soft, quasi-thermal spectrum.  This type of spectrum is consistent with an origin in the inner region of a radiatively efficient accretion flow onto the supermassive black hole (SMBH) residing in the centers of their host galaxies.  The photon energy dependence of the QPE flare amplitude and temporal width (e.g., \citealt{Miniutti+19}; their Fig.~2) also supports an origin for the emission from the innermost radii of a SMBH accretion flow.  Assuming a $10\%$ radiative efficiency, the average mass accreted by the SMBH per eruption to explain the radiated energy is $M_{\rm acc} \sim 10^{-6}M_{\odot}$($10^{-8}M_{\odot}$) in eRO-QPE1 (eRO-QPE2).  Notably, the peak luminosity of the flares can vary by up to an order of magnitude even within a single source \citep{Arcodia+21}.

The stellar masses of the galaxies hosting eRO-QPE1/eRO-QPE2 are relatively low, $M_{\star} \approx 1-4\times 10^{9}M_{\odot}$.  The standard bulge-SMBH mass $M_{\rm bulge}-M_{\bullet}$ relationship points to relatively low-mass SMBHs, with $M_{\bullet} \sim 10^{5}-10^{7} M_{\odot}$, although this relation suffers from large scatter in this range of $M_{\rm bulge} \le M_{\star}$ (e.g., \citealt{Greene+20,Sharma+20}).  This spread also encompasses the range of SMBH masses inferred by X-ray spectral fitting for GSN069 ($M_{\bullet} \approx 4\times 10^{5}M_{\odot}$; \citealt{Miniutti+19}) and RX J1301.9+2747 ($M_{\bullet} \sim 1-3\times 10^{6}M_{\odot}$; \citealt{Giustini+20}). A similar SMBH mass range is needed to match the X-ray luminosities to the range $L_{\rm X} \gtrsim 10^{-2}L_{\rm Edd} \sim 10^{42}(M_{\bullet}/10^{6}M_{\odot})$ erg s$^{-1}$ associated with radiatively efficient accretion (e.g., \citealt{Ho09}).

The host galaxies of GSN 069 and RX J1301.9+2747 exhibit emission lines indicative of active galactic nuclei (AGN) and post-starburst behavior (\citealt{Miniutti+19,Giustini+20}).  However, the nuclei of the {\it eROSITA} QPE hosts appear quiescent, a feature which \citet{Arcodia+21} suggest may make them more representative of the QPE population due to the blind nature of the {\it eROSITA} survey.  Furthermore, no QPE thus far exhibits evidence of optical/UV variability due to reprocessing of the X-ray emission \citep{Miniutti+19,Arcodia+21}, constraining the radial extent of any large-scale accretion flow surrounding the SMBH.  Based on the lack of a detectable narrow-line region, \citet{Arcodia+21} place an upper limit $T_{\rm active} \lesssim 10^{3}-10^{4}$ years on the duration of AGN activity in the QPE hosts.  On the other hand, archival X-ray detections of RX J1301.9+2747 and GSN 069\footnote{We note that GSN 069 exhibits far more long-term X-ray variability than standard AGN, with a first detection in 2010 that is a factor $\approx 240$ brighter than the upper limit from a {\it ROSAT} non-detection in 1994 \citep{Saxton+11, Miniutti+13}.} show these nuclei have been active for at least 18.5 \citep{Giustini+20} and 11 \citep{Miniutti+19} years, respectively.  However, they may not have been generating QPEs this entire time, with {\it XMM} Slew Survey observations of GSN069 and eRO-QPE2 ruling out QPE emission as recently as $\sim$2014 and $\sim$2010 at the level of the present-day quiescent flux (R.~Arcodia, private communication).

\begin{table*}
\centering
\caption{Summary of QPE Properties}
\begin{tabular}{ccccccc}
\hline
$T_{\rm QPE}^{\rm (a)}$ & $\tau_{\rm QPE}^{\rm (b)}$ & $L_{\rm X}^{\rm (c)}$ & $M_{\rm acc}^{\rm (d)}$ & $M_{\bullet}^{\rm (e)}$ & $\tau_{\rm active}^{(f)}$ & $T_{\rm active}^{\rm (g)}$\\
      (hr)  & (hr) & (erg s$^{-1}$) & ($M_{\odot}$) & ($M_{\odot}$) & (yr) & (yr) \\
\hline
$\approx 2-19$ & $\approx 0.4-8$ & $\sim 10^{42}-10^{44}$ & $\gtrsim 10^{-8}-10^{-6}$ & $\sim 10^{5}-10^{6.5}$ & $\lesssim 2$ & $\lesssim 10^{3}-10^{4}$ \\
\end{tabular}
\label{t:summary}
\begin{flushleft}
{\bf Note:} Columns from left to right show: (a) the period separating flares, (b) flare duration, (c) peak X-ray luminosity of flares, (d) inferred accreted mass per flare, (e) SMBH mass, (f) duration of recent QPE activity; and (g) total AGN active duration.
\end{flushleft}
\end{table*}

As summarized in Table \ref{t:summary}, any viable explanation for the QPE phenomenon requires a mechanism capable of abruptly and quasi-periodically feeding the innermost region of a relatively low-mass SMBH (in what is at least sometimes an otherwise quiescent nucleus) with a gaseous mass $\gtrsim 10^{-8}-10^{-6}M_{\odot}$ over a duration $\tau_{\rm QPE} \approx 0.4-8$ hr, recurring regularly every $T_{\rm QPE} \approx 2-19$ hr for at least a period of $\tau_{\rm active} \gtrsim$ 2 years, but associated with longer lived AGN activity of duration 10 yr $\lesssim T_{\rm active} \lesssim 10^{3}-10^{4}$ yr.  

The existence of the QPE phenomena in quiescent galactic nuclei, together with the detailed modeling of the X-ray timing properties \citep{Arcodia+21}, would appear to disfavor explanations that involve instabilities in a long-lived gaseous AGN accretion disk (e.g., \citealt{Miniutti+19}).  Quasi-periodic activity associated with the merger of a binary SMBH (of mass ratio close to unity) is also disfavored by a few arguments (\citealt{Arcodia+21}), in particular the short timescale over which the QPE period would evolve due to gravitational wave-driven orbital evolution.  \citet{Ingram+21} explore the possibility of self-lensing of a massive binary black hole, whereby the ``mini-disk" surrounding one black hole is lensed by the other black hole for an edge-on viewing orientation.  While this model can in principle explain the sharp and symmetric light curve shapes of QPEs, it appears to run into difficulty simultaneously explaining the amplitude and duration of the flares.  Furthermore, lensing should be achromatic, while the QPE duration depends on X-ray photon energy ($\tau_{\rm QPE}$ is smaller in hard X-rays than soft X-rays).

A potentially more promising class of models are those that invoke extreme mass ratio inspiral (EMRI) binaries, since the gravitational inspiral time of an EMRI is considerably longer than for a binary SMBH.  
The steady-state mass transfer rate from a main sequence star onto the SMBH is deeply sub-Eddington \citep{Linial&Sari17} and hence incapable of explaining QPE luminosities (see Eq.~\ref{eq:Lavg} below). \citet{King20} propose that a white dwarf (WD) EMRI on a highly eccentric orbit, which periodically overflows its Roche lobe onto the SMBH, could generate the observed QPEs, a scenario first explored theoretically in \citet{Zalamea+10}.  In Sections \ref{sec:Hills} and  \ref{sec:alternative} we explore this scenario, and the related one involving an ordinary (non-degenerate) star on an eccentric orbit.  We find that the parameter space for forming such short-lived systems is extremely narrow and hence single EMRI explanations are strongly disfavored due to their inability to explain the rate of QPEs inferred from {\it eROSITA}.  

\citet{Sukova+21} employ general relativistic magnetohydrodynamical simulations to explore the impact of an orbiting star embedded in a pre-existing gaseous accretion disk on the black hole accretion rate and disk outflow rate.  They find that quasi-periodic behavior can be induced in the accretion rate by the star, including time-evolution in some models in qualitative agreement with observed QPE light curves.  While promising, the results may be sensitive to several of the simplifying assumptions (such as the neglect of radiative cooling on the disk structure and the use of strong approximations to map the effects of the stellar orbit into a two-dimensional simulation) and the magnetic field evolution in the torus, which depends on the initial magnetic field topology and the grid resolution.  The model also presumes the existence of a radially-extended AGN disk, which though justified for GSN 069 and RX J1301.9+2747, is more questionable for the eROSITA QPEs given their otherwise quiescent host nuclei.

Here, we consider an alternative hypothesis: mass loss due to periodic close interactions between two quasi-circular stellar EMRIs \citep[hereafter MS17]{Metzger&Stone17}.  An EMRI comprised of a main sequence star that inspirals into the SMBH on a nearly circular orbit can undergo Roche lobe overflow (RLOF) and stable mass-transfer onto the SMBH on a radial scale $\sim 1$ AU from the SMBH (e.g., \citealt{King&Done93,Dai&Blandford13,Linial&Sari17}), in analogy to a cataclysmic variable or X-ray binary.  As pointed out by MS17, the timescale for mass-transfer evolution, $\sim 1-10$ Myr, can be comparable to the interval between consecutive circular EMRIs.  As a consequence, the semi-major axis of the more massive EMRI will approach that of the less massive one, leading to periodic strong tidal interactions or even grazing physical collisions between the stars, ultimately destroying one or both bodies.  

MS17 showed that the resulting episodes of gas production, generated each time the EMRIs pass close to one another, could generate QPE-like bursts through quasi-periodic episodes of SMBH accretion.  However, MS17 predicted recurrence times between bursts of $1~{\rm yr} \lesssim T_{\rm QPE} \lesssim 10^4 ~{\rm yr}$, far larger than the observed timescales.  This long delay arose because of their assumption that the two EMRIs occupy distinct orbital planes, a geometry that reduces the interaction probability and increases the interval between consecutive close passages.  

Here, we instead consider the interaction between two {\it co-planar} EMRIs, at least one of which is undergoing RLOF onto the SMBH.  We show that the gravitational force of one EMRI acts to reduce the Hill radius of a RLOFing counterpart, leading to an enhanced mass-transfer rate to the SMBH during the brief periods of closest approach.  We argue that such sequences of flybys can quantitatively account for the timescales, energetics, and rates of the QPE phenomenon.

This paper is organized as follows.  Section \ref{sec:interaction} describes the interaction between coplanar EMRIs, which we compare to QPE observations in Section \ref{sec:observations}.  Section \ref{sec:channels} explores channels for generating circular EMRI pairs.  We discuss our results in Section \ref{sec:discussion} and conclude in Section \ref{sec:conclusions}.

\begin{figure}
    \centering
    \includegraphics[width=0.5\textwidth]{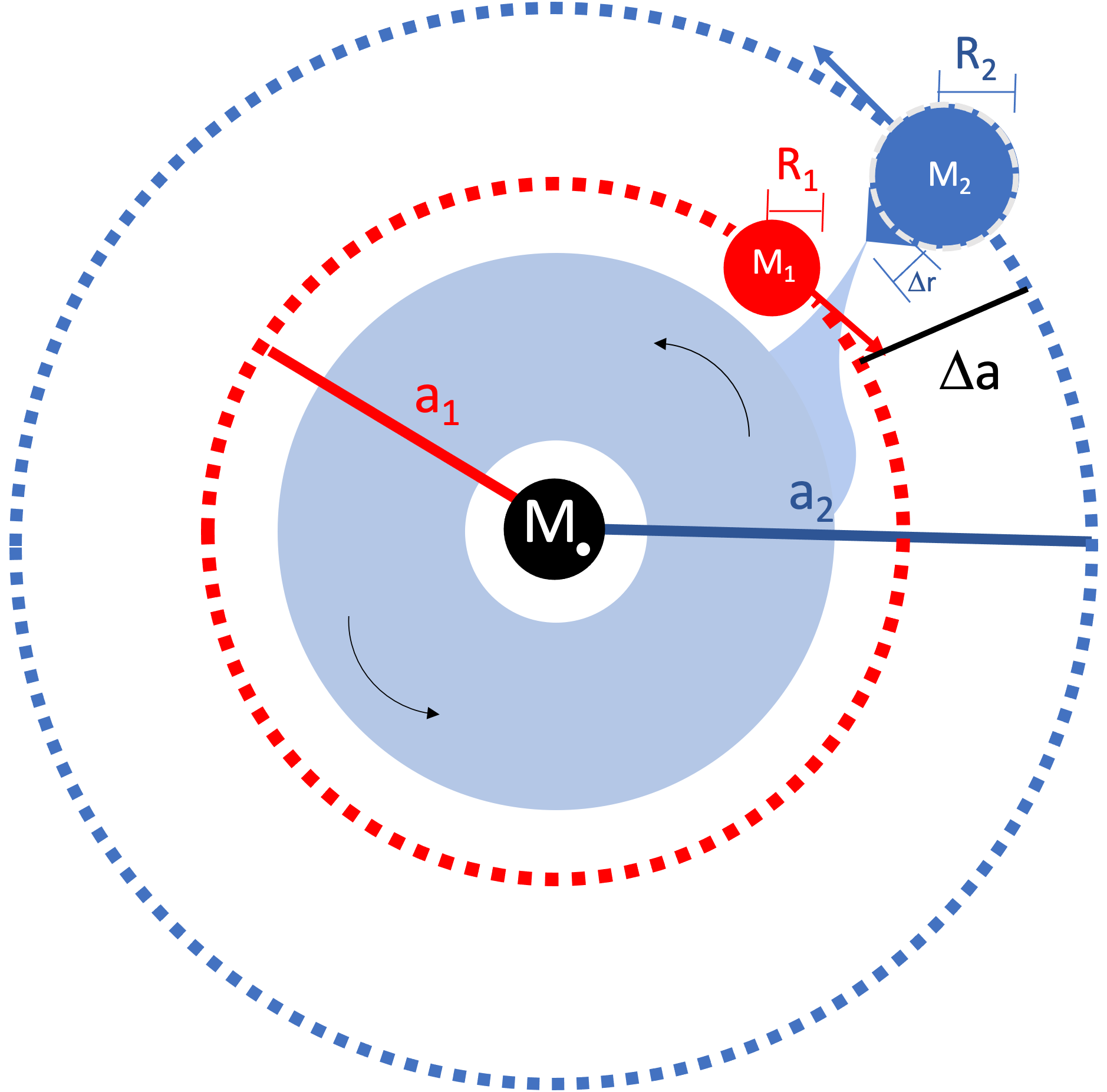}
    \caption{Schematic illustration of the interaction between consecutive coplanar EMRIs, of masses $M_1$, $M_2(M_2 >M_1)$ and radii $R_1$, $R_2 (R_2 > R_1)$, respectively, orbiting a SMBH of mass $M_{\bullet}$ with semi-major axes $a_1$, $a_2$.  In our fiducial scenario, both $M_1$ and $M_2$ fill their Roche lobes, such that each is slowly transferring mass onto the SMBH on a timescale dictated by gravitational wave radiation, until their orbits approach within a separation $\Delta a = a_2-a_1 \lesssim 5R_2$.  In the counter-orbiting case illustrated, the two stars pass within a distance $\sim \Delta a$ twice per (their approximately common) orbital period $\tau_{\rm orb} \sim$ hours.  During this brief flyby, the gravitational influence of $M_1$ acts to shrink the Roche surface of $M_2$ by a distance $\Delta r$, temporarily boosting its mass-loss rate onto the SMBH and generating an observable, accretion-powered X-ray flare.  }
    \label{fig:cartoon}
\end{figure}

\section{Interactions Between Coplanar EMRIs}
\label{sec:interaction}

In this section we estimate the properties of interacting stellar EMRIs and address how they can produce unbound gas through close encounters.  We follow the scenario outlined in MS17, but modified to focus on the case of coplanar orbits.  Figure \ref{fig:cartoon} illustrates the system.

\subsection{EMRI Pairs}

The first EMRI is assumed to be of a star (or brown dwarf or planet) of mass $M_1 = m_1M_{\odot}$ and radius $R_1 = r_1R_{\odot}$.  If $M_1$ is overflowing its Roche lobe onto the SMBH of mass $M_{\bullet} = 10^{6}M_{\bullet,6}M_{\odot}$, then its semi-major axis is given by
\begin{eqnarray}
a_1 &=& R_{\rm RL} \simeq 2.17 R_{1}\left(\frac{M_{\bullet}}{M_{1}}\right)^{1/3} \nonumber \\
&\simeq& 1.0\,{\rm AU}\,M_{\bullet,6}^{1/3}\frac{r_1}{m_1^{1/3}} \approx 1.0{\rm AU}\,\frac{M_{\bullet,6}^{1/3}}{\tilde{\rho}_1^{1/3}},
\label{eq:aRL}
\end{eqnarray}
where $\tilde{\rho}_1 \equiv \rho_1/\rho_{\odot}$ is the mean density $\rho_1$ of $M_1$ normalized to the solar value $\rho_{\odot}$.  The semi-major axis must also exceed that of the innermost stable circular orbit (ISCO),
\be
R_{\rm ISCO} \approx 0.06\,{\rm AU}(R_{\rm ISCO}/6R_{\rm g})M_{\bullet,6}
\label{eq:RISCO}
\ee
where the gravitational radius $R_{\rm g} \equiv GM_{\bullet}/c^{2}$ and $R_{\rm ISCO}/R_{\rm g}$ varies from 1 to 9 as the dimensionless SMBH spin $a_{\bullet}$ varies from $+1$ to $-1$.  We see that $R_{\rm RL} > R_{\rm ISCO}$ for SMBH masses in the range estimated from QPE host galaxies ($M_{\bullet} \lesssim 10^{6}M_{\odot}$), for all physically allowed values of $M_{1}$, $R_{1}$ corresponding to brown dwarfs or non-degenerate stars.

The first EMRI $M_1$ undergoes RLOF evolution on the timescale set by gravitational wave radiation (e.g., MS17),
\begin{equation}
\tau_{\rm GW} \approx 1.3\times 10^{6}\,{\rm yr} \,\,\chi M_{\bullet,6}^{-2/3}m_1^{-1}\tilde{\rho}_1^{-4/3}.
\label{eq:tauGW}
\end{equation}
The dimensionless factor $\chi = 1$ in the case of free inspiral and $\chi = 3/(3p-1) \approx 2-4$ if $M_1$ is undergoing RLOF, where $R_{\star} \propto M_{\star}^{p}$ and the given range of $\chi$ corresponds to $p \approx 0.6-0.8$ for a range of stellar masses and thermal states (e.g., \citealt{Linial&Sari17}; MS17).  Assuming stable mass transfer, this results in a mass-accretion rate $\langle \dot{M}\rangle \sim M_1/\tau_{\rm GW}$ and corresponding accretion luminosity,
\be \langle L \rangle \approx 0.1\langle \dot{M} \rangle c^{2} \approx 4\times 10^{39}{\rm erg\,s^{-1}}\,\,\chi^{-1}M_{\bullet,6}^{2/3}m_1^{2}\tilde{\rho}_1^{4/3}.
\label{eq:Lavg}
\ee
This is too small to explain time-averaged QPE luminosities by several orders of magnitude, demonstrating why QPE models which invoke single EMRIs would require unstable (runaway) mass-transfer (which, however, has its own challenges; Section \ref{sec:alternative}).

The second EMRI is a star of mass $M_2 = m_2 M_{\odot}$ and radius $R_2 = r_2 R_{\odot}$ on an orbit of semi-major axis $a_2$.  We assume that both EMRIs have nearly circularized their orbits due to energy loss via gravitational wave (GW) emission.  A strong interaction between two consecutive EMRIs will only occur if their orbits approach one another because the rate of gravitational wave-driven orbital decay of $M_2$ is faster than that of $M_1$.  We thus require $M_2 \gtrsim M_1$ for an interaction.  If both $M_1$ and $M_2$ are filling their Roche radii, then they must possess roughly equal mean densities due to their common semi-major axes near the point of strongest interaction, i.e. $M_1/R_1^{3} \approx M_2/R_2^3$ and hence we also require $R_2 \gtrsim R_1$.  

\subsection{Condition for Close Interactions}

Once the orbits of the two EMRIs approach within a separation $\Delta a \equiv a_2-a_1$ of several stellar radii, strong tidal interactions occur between them.  At this point the EMRIs share a roughly common semi-major axis $a_1 \simeq a_2 = a$ and orbital period,
\begin{eqnarray}
T_{\rm orb} &\simeq& 2\pi \left(\frac{a^{3}}{GM_{\bullet}}\right)^{1/2} \approx 8.8\,{\rm hr}\,\,\tilde{\rho}^{-1/2}\left(\frac{a}{R_{\rm RL}}\right)^{3/2},
\label{eq:tauorb}
\end{eqnarray} 
where $\tilde{\rho}$ is the mean density of either star with respective Roche radius $R_{\rm RL}$ (Eq.~\ref{eq:aRL}).

To simplify the analysis below, due to the slower evolution of $M_1$ we approximate its orbit as being fixed during its interaction with $M_2$.  The number of orbits required for $M_2$ to migrate inward radially, via gravitational wave emission, by a distance $\delta a \ll a_2$ is given by 
\begin{eqnarray}
N_{\rm GW} &\sim& \frac{\tau_{\rm GW}}{T_{\rm orb}}\left(\frac{\delta a}{a_2}\right) \nonumber \\
 &\approx& 6\times 10^{6}\frac{\chi}{M_{\bullet,6}}\frac{r_2}{m_2 \tilde{\rho}_2^{1/2}}\left(\frac{a}{R_{\rm RL}}\right)^{3/2}.
\label{eq:NGW},
\end{eqnarray}
where for $\tau_{\rm GW}$ we use Eq.~(\ref{eq:tauGW}) replacing $M_1$ with $M_2$.  

Each close flyby will result in the removal of mass from one or both stars and an accretion-powered flare, such that the QPE recurrence time $T_{\rm QPE}$ is roughly the time between flybys, $T_{\rm fly}$ (however, see Section \ref{sec:nonperiodic}).  The specific mechanism of the mass removal is described below.  There are two cases to consider, depending on whether both EMRIs are orbiting in the same direction (``co-orbiting'' case) or in opposite directions (``counter-orbiting'' case).  

In the counter-orbiting case, close passages occur twice per orbital period (Eq.~\ref{eq:tauorb}),
\be
T_{\rm fly} \approx \frac{T_{\rm orb}}{2} \approx 4.4\,{\rm hr}\,\,\tilde{\rho}^{-1/2},  \text{Counter-orbiting.}
\label{eq:taucoll1}
\ee
where $\tilde{\rho}$ is the mean density of the star or stars undergoing RLOF (in our fiducial scenario, at least $M_2$).  

\begin{figure*}
    \centering
    \includegraphics[width=1.0\textwidth]{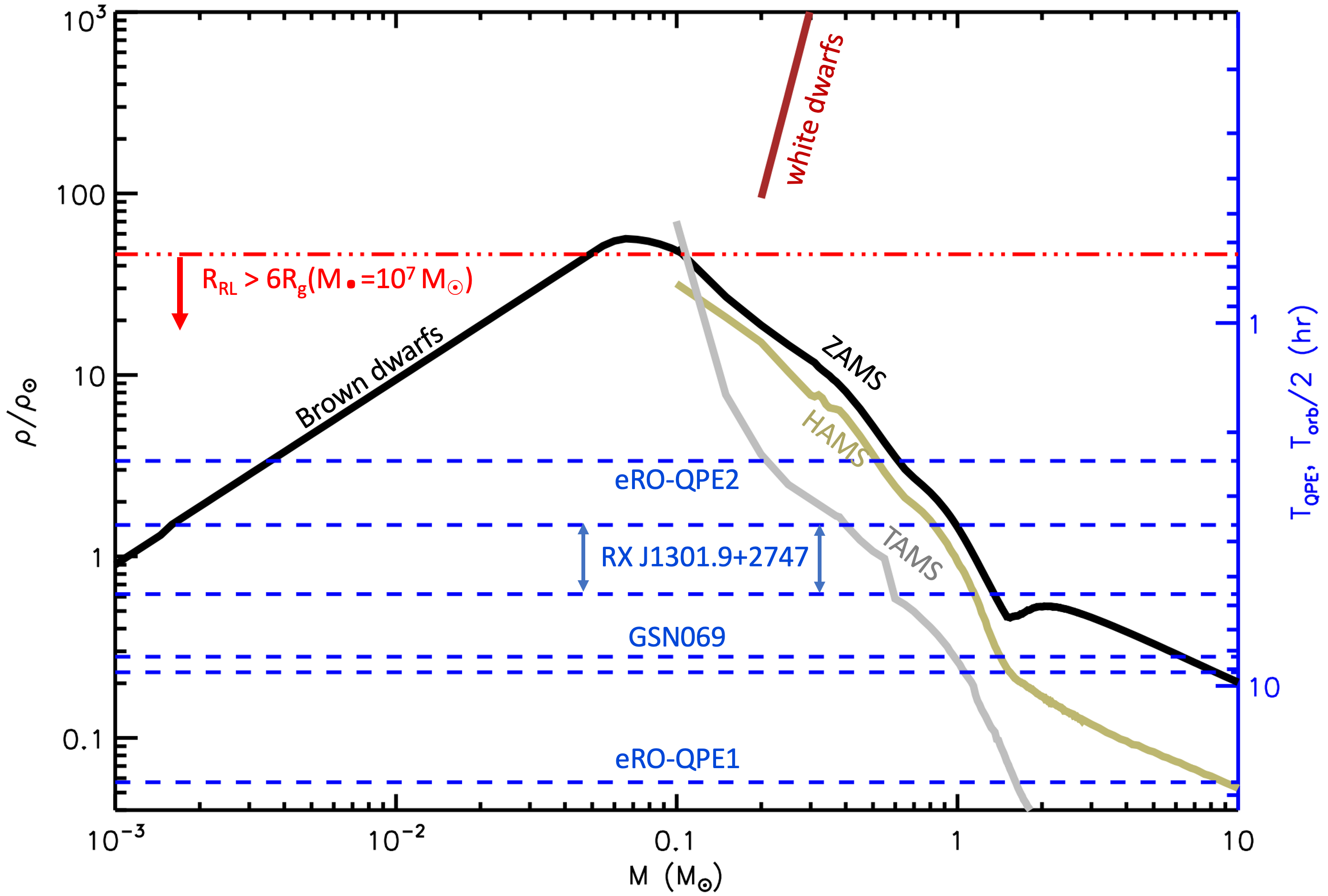}
    \caption{Mean density $\rho$ of a stellar EMRI (normalized to the solar value $\rho_{\odot}$) undergoing RLOF onto the SMBH as a function of the EMRI mass $M$.  A brown line shows WDs, while a black line shows tracks corresponding to Solar metallicity stars on the zero-age main-sequence stars (ZAMS), and brown dwarfs/gas giant planets (making the approximation of a constant radius of $0.1 R_{\odot}$; \citealt{Chabrier+09}).  Olive and gray lines show the half-age main sequence (HAMS) and terminal-age main sequence (TAMS), respectively (all main sequence tracks are calculated from the MIST database of MESA stellar evolution models; \citealt{Dotter16,Choi+16,Paxton+11,Paxton+13,Paxton+15,Paxton+19}).  For comparison we show the observed QPE periods associated with eRO-QPE1/eRO-QPE2 \citep{Arcodia+21}, RX J1301.9+2747 \citep{Giustini+20}, GSN 069 \citep{Miniutti+19} in the case of QPEs arising from close interactions between counter-orbiting EMRIs (Eq.~\ref{eq:taucoll1}).  Insofar that that the ZAMS/HAMS/TAMS lines represent the maximum density of a star of a given mass and nuclear evolutionary state (i.e., not accounting for puffing up of the star due to adiabatic mass-loss; \citealt{Linial&Sari17}), the intersection of these lines with observed QPE periods represents a maximum on the mass of the RLOFing star ($M_2$) responsible for generating the observed X-ray flares.}  
    \label{fig:density}
\end{figure*}

Figure \ref{fig:density} shows the mean density of stars in different evolutionary stages (WDs, brown dwarfs, and stars at different phases of the main sequence) as a function of their mass, compared to the minimum density compatible with the observed values of $T_{\rm QPE} = T_{\rm fly}$ according to Eq.~(\ref{eq:taucoll1}) assuming $a = R_{\rm RL}$.  For example, eRO-QPE1(eRO-QPE2) require $\rho/\rho_{\odot} \gtrsim 0.06(2.7)$ to match the observed eruption periods $T_{\rm QPE} \approx 18.5(2.7)$ hr.  eRO-QPE2 is consistent with a brown dwarf/planet or ZAMS stars of mass $3\times 10^{-3} \lesssim M_1 \lesssim 0.7M_{\odot}$.  The longer period of eRO-QPE1 is not compatible with ZAMS star undergoing RLOF, but is compatible with an evolved star of mass $\gtrsim 2M_{\odot}$.  

The ZAMS/HAMS/TAMS stellar properties shown in Fig.~\ref{fig:density} assume thermal equilibrium, which is not a good approximation when the stars are losing mass at a high rate (e.g., \citealt{Linial&Sari17}; see discussion at the end of Appendix \ref{sec:flyby}).  Insofar that thermal timescale mass-loss will cause a star to inflate, the lines in Fig.~\ref{fig:density} represent an upper limit on the inferred density of the RLOFing star (lower limit on $T_{\rm QPE}$) at a given stellar mass.  

Next consider the co-orbiting case.  Here, the inner star $M_1$ must ``chase'' the outer one $M_2$ due to its slightly shorter orbital period $\Delta T_{\rm orb} \ll T_{\rm orb}$, as results from their small semi-major axis difference, $\Delta a \ll a_1, a_2$.  The greater number of orbits required for a close passage in this case, $N_{\rm fly} \approx T_{\rm orb}/|\Delta T_{\rm orb}| \approx (2/3)(a/\Delta a) \gg 1$, results in a larger time interval between collisions,
\be
T_{\rm fly} \approx N_{\rm fly}T_{\rm orb} \approx 10.8\,{\rm d}\,\frac{M_{\bullet,6}^{1/3}}{m_2^{1/3}\tilde{\rho}^{1/2}}\left(\frac{\Delta a}{5R_2}\right)^{-1}, \text{Co-orbiting,}
\label{eq:taucoll2}
\ee
where we have assumed $M_2$ is overflowing its Roche Lobe and $\Delta a$ is normalized to a characteristic value $\sim 5R_2$ necessary for a strong interaction (Eq.~\ref{eq:deltamdest} below).  The large value of $T_{\rm fly}$ in the co-orbiting case is challenging to reconcile with the short observed QPE periods $T_{\rm QPE} \sim$ hours, unless the colliding stars are WDs ($r_{2} \sim 0.01$; $\rho \gtrsim 10^{5}$ g cm$^{-3}$).  For this reason we favor the counter-orbiting case.  However, our results to follow would apply equally to the co-orbiting case, and the latter may be relevant for longer-period AGN variability (Section \ref{sec:ASASSN}).

During the interval between close encounters, $T_{\rm fly}$, the orbital separation $\Delta a$ decreases due to the gravitational wave inspiral of $M_2$ by an amount $\Delta a_{\rm GW}$, obtained by setting $N_{\rm GW} = N_{\rm fly} = 1/2$ and $N_{\rm GW} = N_{\rm fly} \simeq (2/3)(a/\Delta a)$ in the counter-orbiting and co-orbiting cases, respectively:
\begin{eqnarray}
&&\frac{\Delta a_{\rm GW}}{R_{2}} \approx 8\times 10^{-8}\frac{r_2^{2}}{\chi}M_{\bullet,6} \tilde{\rho}^{3/2},\,\,\text{Counter-orbiting}, \nonumber \\
&\approx& 5\times 10^{-6} \frac{r_2}{\chi}M_{\bullet,6}^{4/3}\tilde{\rho}^{7/6}\left(\frac{\Delta a}{5R_2}\right)^{-1},\,\,\text{Co-orbiting},
\label{eq:deltab}
\end{eqnarray}
where we have used Eq.~(\ref{eq:NGW}) with $\delta a = \Delta a_{\rm GW}$ and have assumed both stars are overflowing their Roche lobes.  For stellar parameters $\{\rho \sim 0.1-10$ g cm$^{-3}$, $r_2 \sim 0.1-1\}$ and $\{\rho \gtrsim 10^{5}$ g cm$^{-3}$, $r_2 \sim 0.01\}$ necessary to match the QPE timescales in the counter-orbiting and co-orbiting cases, respectively, we have $\Delta a_{\rm GW}/R_2 \ll 1$.  The two stars will thus be subject to many strong flybys prior to any direct contact between their surfaces.

\subsection{Mass Loss from Flybys}

A close passage between $M_1$ and $M_2$ can generate mass loss from one or both stars exceeding their rate of steady mass-transfer onto the SMBH.  Mass loss can in principle arise either from a direct physical collision between the stars (``hydrodynamical'' mass-loss), or as the result of tidal forces impacting the rate of mass-transfer onto the SMBH (``tidal'' mass-loss).  In both cases the more compact lower-mass star $M_1$ will preferentially remove mass from the more dilute outer layers of $M_2$.  For this reason and others related to the geometry of the Roche surface (see below), the bulk of this discussion focuses on mass loss from $M_2$.  Furthermore, we focus on tidal instead of hydrodynamical mass-loss because: (1) as we show below, it becomes significant once $\Delta a$ shrinks to a few stellar radii; (2) many such close flybys occur before the first physical collision (Eq.~\ref{eq:deltab}).  The latter point contrasts with the non-coplanar case, for which many more orbits separate the close encounters and physical collisions are more relevant (MS17).

As we show in Appendix \ref{sec:flyby}, the gravitational influence of a close passage from $M_1$ is to briefly shrink the Hill radius $r_{\rm H}$ of $M_2$, according to:
\be
\frac{r_{\rm H}}{r_{\rm H,0}} \equiv 1-\epsilon; \,\,\,\, \epsilon \simeq \frac{M_{1}}{3 M_2}\left(\frac{\Delta a}{r_{\rm H,0}}\right)^{-2},
\ee
where $r_{\rm H,0} \simeq (M_2/3M_{\bullet})^{1/3}$ is the usual (unperturbed) Hill radius (Eq.~\ref{eq:aRL}).  

Insofar that $r_{\rm H,0} \simeq R_2$ if $M_2$ is filling its Roche lobe and losing mass through the inner Lagrange point $L_1$, then the close passage of $M_1$ causes the Roche surface of $R_2$ to penetrate below its photosphere by an additional factor $\Delta r \simeq \epsilon R_{2}$.  To the extent that $\Delta r$ exceeds the atmosphere scale-height $H \sim (10^{-4}-10^{-3})R_2$ near the photosphere of $M_2$, this increases its mass-loss rate through $L_1$ by a large factor for the brief time interval $\tau_{\rm fly} \sim (\Delta a/a)T_{\rm QPE}$ the two stars spend close to each other.

In Appendix \ref{sec:flyby} we estimate the mass-loss $\Delta m_{\rm fly}$ from $M_2$ per close passage, following the formalism of \citet{Ginzburg&Quataert21}.  We find (Eq.~\ref{eq:deltam_app})
\begin{eqnarray}
&&\Delta m_{\rm fly} \sim 5\times 10^{-8}M_{\odot}\tilde{\kappa}^{-1}\frac{m_1^{4.5}}{r_2^{3.5}}\times \nonumber \\
&&\frac{m_2^{1/3}}{M_{\bullet,6}^{1/3}}\left(\frac{T_{\rm QPE}}{10\,{\rm hr}}\right)\left(\frac{T_{\rm eff}}{10^{4}{\rm K}}\right)^{-4}\left(\frac{\Delta a}{5 R_2}\right)^{-8},
\label{eq:deltam}
\end{eqnarray}
where $\tilde{\kappa}$ is the photosphere opacity (normalized to the electron scattering opacity) and $T_{\rm eff} \sim 10^{4}$ K the surface temperature, where we have assumed an $n = 3$ polytrope for the outer envelope structure.  These surface properties are expected due to the strong influence of irradiation of the star by the luminous SMBH accretion flow, which usually overwhelms its internal nuclear luminosity (Appendix \ref{sec:flyby}).  

The total mass loss from $M_2$ during the time the two stars spend separated by any distance $\sim \Delta a$ is given by
\begin{eqnarray}
\frac{\Delta m}{M_2} &\sim& N_{\rm GW}(\Delta a)\frac{\Delta m_{\rm fly}}{M_2} \nonumber \\
&\sim& 1.2 M_{\bullet,6}^{-4/3}\frac{\chi }{\tilde{\kappa}}\frac{m_1^{4.5}}{m_2^{2} r_2^{1.5}}\left(\frac{T_{\rm QPE}}{10\,{\rm hr}}\right)\left(\frac{T_{\rm eff}}{10^{4}{\rm K}}\right)^{-4}\left(\frac{\Delta a}{5 R_2}\right)^{-7}
\end{eqnarray}
where we have used Eq.~(\ref{eq:NGW}) for $N_{\rm GW}$ with $a = R_{\rm RL}$ and $\delta a = \Delta a$.  

We thus see that $M_2$ will be completely destroyed ($\Delta m \gtrsim M_2$) once gravitational wave radiation reduces the orbital separation $\Delta a$ below a critical value 
\begin{eqnarray}
 \frac{\Delta a_{\rm dest}}{R_2} &\approx& 
 5 M_{\bullet,6}^{-4/21}\frac{\chi^{1/7}}{\tilde{\kappa}^{1/7}}\frac{m_1^{9/14}}{r_2^{3/28}m_2^{2/7}} \left(\frac{T_{\rm QPE}}{10\,{\rm hr}}\right)^{1/7}\left(\frac{T_{\rm eff}}{10^{4}\,{\rm K}}\right)^{-4/7}, \nonumber \\
 \label{eq:deltaa_dest}
\end{eqnarray}
which we note is a weak function of the relevant parameters.  The destruction of $M_2$ will occur gradually, over a timescale,
\begin{eqnarray}
t_{\rm dest} 
 &\sim& N_{\rm GW}(\Delta a_{\rm dest})T_{\rm orb} \nonumber \\
 &\approx& 3\times 10^{4}\,{\rm yr}\frac{\chi}{M_{\bullet,6}}\frac{r_2^{7/2}}{m_2^{11/6}}\left(\frac{\Delta a_{\rm dest}}{5 R_2}\right).
\label{eq:tdest}
\end{eqnarray}  
Most of the total mass of $M_2$ accreted by the SMBH will therefore occur when the per-flyby mass-loss $\Delta m_{\rm fly}$ is near the critical value
\begin{eqnarray}
&\Delta m_{\rm dest}& \sim M_2\frac{T_{\rm QPE}}{t_{\rm dest}}  \nonumber \\
&\sim& 4\times 10^{-8}M_{\odot} \frac{M_{\bullet,6}}{\chi}\frac{m_2^{17/6}}{r_2^{7/2}}\left(\frac{T_{\rm QPE}}{10\,{\rm hr}}\right)\left(\frac{\Delta a_{\rm dest}}{5 R_2}\right)^{-1}. \nonumber \\
\label{eq:deltamdest}
\end{eqnarray}
We thus find that $\Delta m_{\rm dest}$ and $t_{\rm dest}$ are constrained to lie within a couple orders of magnitude of $\sim 10^{-8}M_{\odot}$ and $\sim 10^{4}$ yr, respectively (for the allowed ranges of $m_2$ and $M_{\bullet}$).

The above expressions refer to mass-loss from the inner $L_1$ point of $M_2$ due to tidal interactions with $M_1$.  In the case when both stars are undergoing RLOF, $M_1$ can also experience enhanced mass loss through its the outer $L_2$ Lagrange point due to the gravitational force of $M_2$.  However, because of the significant radial separation between the unperturbed $L_1$ and $L_2$ points $\Delta R_{\rm L1,L2}/R_1 \approx (2/3)(M_{1}/M_{\bullet})^{1/3} \sim 10^{-2}$ \citep{Linial&Sari17} relative to the photo-sphere scale-height of $M_2$ (to which the mass-loss rate is extremely sensitive), mass-loss from $M_1$ during the flyby will generally be smaller than that from $M_2$.    

The estimates presented so far assume perfectly circular orbits.  While our scenario invokes quasi-circular EMRI orbits, some residual eccentricity $e \ll 1$ may be present (Eq.~\ref{eq:emax}).  This residual eccentricity can modulate the QPE peak luminosity, $L_{\rm peak}$, in an observable way.  Since $L_{\rm peak} \propto \Delta m_{\rm fly} \propto \Delta a^{-8} \propto (1-e)^{-8}$ (Eq.~\ref{eq:deltam}), even a residual eccentricity of $e\sim 10^{-2}$ ($e\sim 10^{-3}$) suffices to change $L_{\rm peak}$ by a factor of 10 (by a factor of 2), possibly contributing to the large observed variation in QPE amplitudes within a single source.  See Appendix \ref{sec:flyby} for more details.

\section{Comparison to QPE Observations}
\label{sec:observations}

Using results from the previous section, we now examine whether tidally-interacting EMRIs can account for the timescales, energetics, and active durations of QPEs.  The formation channels for coplanar EMRIs are addressed in the next section.

\subsection{QPE Period and Flare Duration}
\label{sec:timescale}

To zeroth order, the QPE period equals the interval between flybys, i.e.~$T_{\rm QPE} = T_{\rm fly}$ (Eqs.~\ref{eq:taucoll1}, \ref{eq:taucoll2}).  The gaseous disk generated by the stripped mass will accrete onto the SMBH, powering X-ray emission, nominally on the viscous time, $\tau_{\rm visc}$, at the circularization radius, $r_{\rm circ}$.  Associating the viscous time with the QPE flare duration, 
\begin{eqnarray}
&&\tau_{\rm QPE} \sim \tau_{\rm visc} \sim \left.\frac{r^{2}}{\nu}\right|_{r_{\rm circ}} \approx \frac{1}{\alpha}\frac{T_{\rm orb}}{2\pi}\left(\frac{h}{r}\right)^{-2} \nonumber \nonumber \\
&\approx&  \frac{3.2}{\alpha_{0.1}}T_{\rm QPE}\left(\frac{h}{r}\right)^{-2}\left(\frac{r_{\rm circ}}{R_{\rm RL}}\right)^{3/2}, \text{Counter-orbiting} \nonumber \\
&\approx& \frac{0.054}{\alpha_{0.1}}T_{\rm QPE}\left(\frac{h}{r}\right)^{-2}\left(\frac{\Delta a}{5R_{2}}\right)\frac{m_1^{1/3}}{M_{\bullet,6}^{1/3}}\left(\frac{r_{\rm circ}}{R_{\rm RL}}\right)^{3/2}, \text{Co-orbiting} \nonumber \\ 
\label{eq:tauQPE}
\end{eqnarray}
where $\nu = \alpha c_{\rm s}h$ is the kinematic viscosity, $h$ the vertical aspect ratio, $c_{\rm s} = h\Omega_K$ the sound speed, $\Omega_{K} = (GM_{\bullet}/r^{3})^{1/2}$, and $\alpha = 0.1\alpha_{0.1}$ the viscosity parameter.  

The duty cycle $\tau_{\rm QPE}/T_{\rm QPE} \sim 0.1-0.4$ inferred from observations of QPEs (e.g., Table \ref{t:summary}) is difficult to satisfy in the counter-orbiting case based on Eq.~(\ref{eq:tauQPE}) if $r_{\rm circ} \sim R_{\rm RL}$ and $h/r \ll 1$.   However, note that: (1) the accretion rate from an initially thin ring of material typically peaks at $\sim 1/10$ of $t_{\rm visc}$ as measured at the ring radius (e.g., \citealt{Pringle81}); (2) disk material formed from the collision will be hot and may find itself in a slim-disk like state (e.g., \citealt{Abramowicz+88}) with $h/r \sim 1$, if the accretion luminosity is indeed approaching the Eddington value, $L_{\rm Edd} \sim 10^{44}M_{\bullet,6}$ erg s$^{-1}$, as may be achieved depending on the SMBH mass; (3) if both stars lose significant mass from the interaction, then due to the opposing specific angular momenta of the counter-orbiting stellar orbits, the disk that forms from the mixture of debris will circularize at radii $r_{\rm circ} < R_{\rm RL}$; (4) systems with $T_{\rm QPE} > \tau_{\rm QPE}$ would not exhibit strong X-ray periodicity and hence would observationally selected against in QPE searches.  

\subsection{QPE Activity Window: SMBH Spin-Induced Nodal Precession}
\label{sec:precession}

We have seen that achieving a match between theoretical and observed values of $T_{\rm QPE}$ is only possible if both stars share a common orbital plane.  However, even if this is true at one moment in time, it may not be true later, due to the effect of nodal precession from the SMBH spin.  

If we assume that both stars are misaligned from the SMBH equatorial plane by an angle $I$, then the maximum distance between the two orbits is $d_{\rm max} \approx a \sin I \sin( \Omega_1 - \Omega_2$), where $\Omega_1$ and $\Omega_2$ are the nodal angles of each orbit with respect to a reference direction in the SMBH equatorial plane, and the approximate equality here reflects the assumption that $\Omega_1 - \Omega_2 \ll 1$ (i.e., the orbits are nearly coplanar).  At leading post-Newtonian order, nodal precession is driven by Lense-Thirring frame dragging, with the nodal shift per orbit for a circular orbit given by
\begin{equation}
    \Delta \Omega = 4\pi \chi_\bullet \left(\frac{a}{R_{\rm g}} \right)^{-3/2}
\end{equation}
where $0 \le \chi_\bullet \le 1$ is the dimensionless spin magnitude of the SMBH \citep{Merritt+10}.  Differential nodal precession will cause initially coplanar orbits to precess into a 3D configuration, so long as their semimajor axes $a_1$ and $a_2 = a_1 + \Delta a$ differ slightly.  After a time $t$, two initially co-aligned orbits will achieve a nodal separation
\begin{eqnarray}
    \Omega_1 - \Omega_2 &=&  \left(\frac{\Delta \Omega_1}{T_{\rm orb,1}} - \frac{\Delta \Omega_2}{T_{\rm orb,2}}\right)t
    \underset{\Delta a \ll 1}\approx  6 \chi_\bullet \frac{G^2M_\bullet^2 t}{c^3 a^3} \frac{\Delta a}{a},  \nonumber \\
\end{eqnarray}
where $T_{\rm orb,1}$ and $T_{\rm orb,2}$ are the orbital periods of $M_1$ and $M_2$, respectively.

The assumption of coplanarity will break down (and QPEs will turn off, for a time) once $d_{\rm max} \gtrsim \Delta a_{\rm dest} \sim 5 R_2$ (Eq.~\ref{eq:deltaa_dest}).  In the small precession limit, $\Omega_1 - \Omega_2 \approx \sin(\Omega_1 - \Omega_2)$, initially coplanar orbits will cease producing QPEs after a time 
\begin{eqnarray}
    \label{eq:Tprec}
    T_{\rm prec} &\approx&  \frac{1}{6}\frac{1}{\chi_{\bullet}\sin I}\frac{c^{3}a^{3}}{G^{2}M_{\bullet}^{2}} 
\approx  100\,{\rm days}\,M_{\bullet,6}^{-2}\left(\frac{0.1}{\chi_{\bullet}I}\right)\left(\frac{a}{\rm AU}\right)^{3} \nonumber \\
&\approx& 100\,{\rm days}\,M_{\bullet,6}^{-1}\tilde{\rho}^{-1}\left(\frac{0.1}{\chi_{\bullet}I}\right),
  \end{eqnarray}
  where in the final line we have taken $a = R_{\rm RL}$ (Eq.~\ref{eq:aRL}).  
  
Precession can thus lead to a long-term modulation in the QPE activity on the timescale $T_{\rm prec}$.  For modest values of the SMBH spin and/or inclination, Eq.~\ref{eq:Tprec} shows that $T_{\rm prec}$ is approaching the active timescale of known QPE systems - for example, eRO-QPE1 has been seen to persist for at least $T_{\rm active} \gtrsim 400$ days (Arcodia, private communication), although the low inferred stellar density in this case $\tilde{\rho} \lesssim 0.1$ (Fig.~\ref{fig:density}) acts to increase $T_{\rm prec}$.  On the other hand, observations of GSN069 in 1990 \citep{Shu+18,Miniutti+19} and eRO-QPE2 in 2014 \citep{Arcodia+21} rule out QPE emission at the level of the present-day quiescent flux, consistent with a scenario in which precession recently brought these systems into alignment.  

If future observations demonstrate that some QPEs do not turn off on the timescale $\sim T_{\rm prec}$, then they must arise not merely from coplanar EMRI pairs, but from pairs that lie within the SMBH equatorial plane, at least to within an angle $\sim \Delta a_{\rm dest}/a \sim 5R_2 / a \sim 10^{-2}$.  This has implications for the required EMRI rate in different formation channels (Section \ref{sec:rates}).  

Finally, we note that a baseline quiescent level of SMBH X-ray accretion activity would be expected, even at times when the EMRI orbits are not aligned to enable strong periodic tidal stripping and QPE emission; this is due to the elevated mass-transfer rate of $M_2$ which results from it being over-inflated as a result of the most recent period of tidally-enhanced mass-loss (Appendix \ref{sec:flyby}).

\subsection{Accreted Mass and Active Duration}
\label{sec:duration}

The maximum mass stripped per EMRI flyby is given by $\Delta m_{\rm dest}$ (Eq.~\ref{eq:deltamdest}).  The predicted range of $\Delta m_{\rm dest} \sim 10^{-9}-10^{-7}M_{\odot}$ is broadly consistent with the radiated X-ray energy of QPE flares \citep{Arcodia+21}, and is relatively insensitive to the free parameters (e.g. $m_1, m_2, r_2, T_{\rm eff}, \chi$).  
The predicted positive correlation between $\Delta m_{\rm dest} \propto T_{\rm QPE}$ is consistent with the flare luminosity of eRO-QPE1 ($T_{\rm QPE} \approx 18.5$ hr) being an order of magnitude higher than that of eRO-QPE2 ($T_{\rm QPE} \approx 2.4$ hr). Indeed, this trend of increasing peak luminosity QPE period is shared by all four known QPEs \citep{Arcodia+21}.

The maximum duration of SMBH activity from EMRI tidal stripping is set by the timescale for strong encounters to destroy one or both stars.  The destruction of $M_2$ occurs on the timescale $t_{\rm dest} \sim 10^{3}-10^{5}$ yr (Eq.~\ref{eq:tdest}), consistent with the upper limit on the AGN activity age $T_{\rm active} \lesssim 10^{3}-10^{4}$ yr in eRO-QPE1 and eRO-QPE2 based on the lack of narrow-line emission from the nuclei of their host galaxies (\citealt{Arcodia+21}; Section \ref{sec:galaxies}).
Although the destruction of $M_2$ can take place over tens of thousands of years or longer, we note that flyby-powered QPEs may not be visible throughout this entire interval due to precession of the EMRI orbital planes by the SMBH spin (on a timescale of $T_{\rm prec} \sim$ months$-$years; Section \ref{sec:precession}).

Another effect that could potentially reduce the lifetime of the interacting EMRI system is ablation of the stars due to interaction with the gaseous accretion disk.  Ablation will be particularly strong in the counter-orbiting case in which $M_1$ orbits in the opposite direction of the disk seeded by mass-loss from $M_2$.
In Appendix \ref{sec:ablation} we estimate the ablation timescale of the star, $t_{\rm abl}$ (Eq.~\ref{eq:tabl}).  For typical gaseous disk properties (e.g., $h/r \gtrsim 0.1$, $L_{\rm X} \sim 10^{42}$ erg s$^{-1}$) we find that $t_{\rm abl} \gtrsim t_{\rm dest} \sim 10^{3}-10^{5}$ yr (Eq.~\ref{eq:tdest}) and hence gas ablation is unlikely to destroy the stars faster than their own self-interaction.  Nevertheless, $t_{\rm abl}$ may be comparable to the lifetime of a pre-existing AGN or the radial migration time of the two EMRIs to their interaction radius (Section \ref{sec:channels}).  In the case of a pre-existing AGN, destruction of the stellar EMRI could in principle also occur due to interaction with a relativistic jet from the SMBH (e.g., \citealt{Zajacek+20}) case when the EMRI orbit is misaligned with the plane of the AGN disk and crosses the jet axis.

\subsection{Deviations from Periodic Behavior}
\label{sec:nonperiodic}

\begin{figure}
    \centering
    \includegraphics[width=0.47\textwidth]{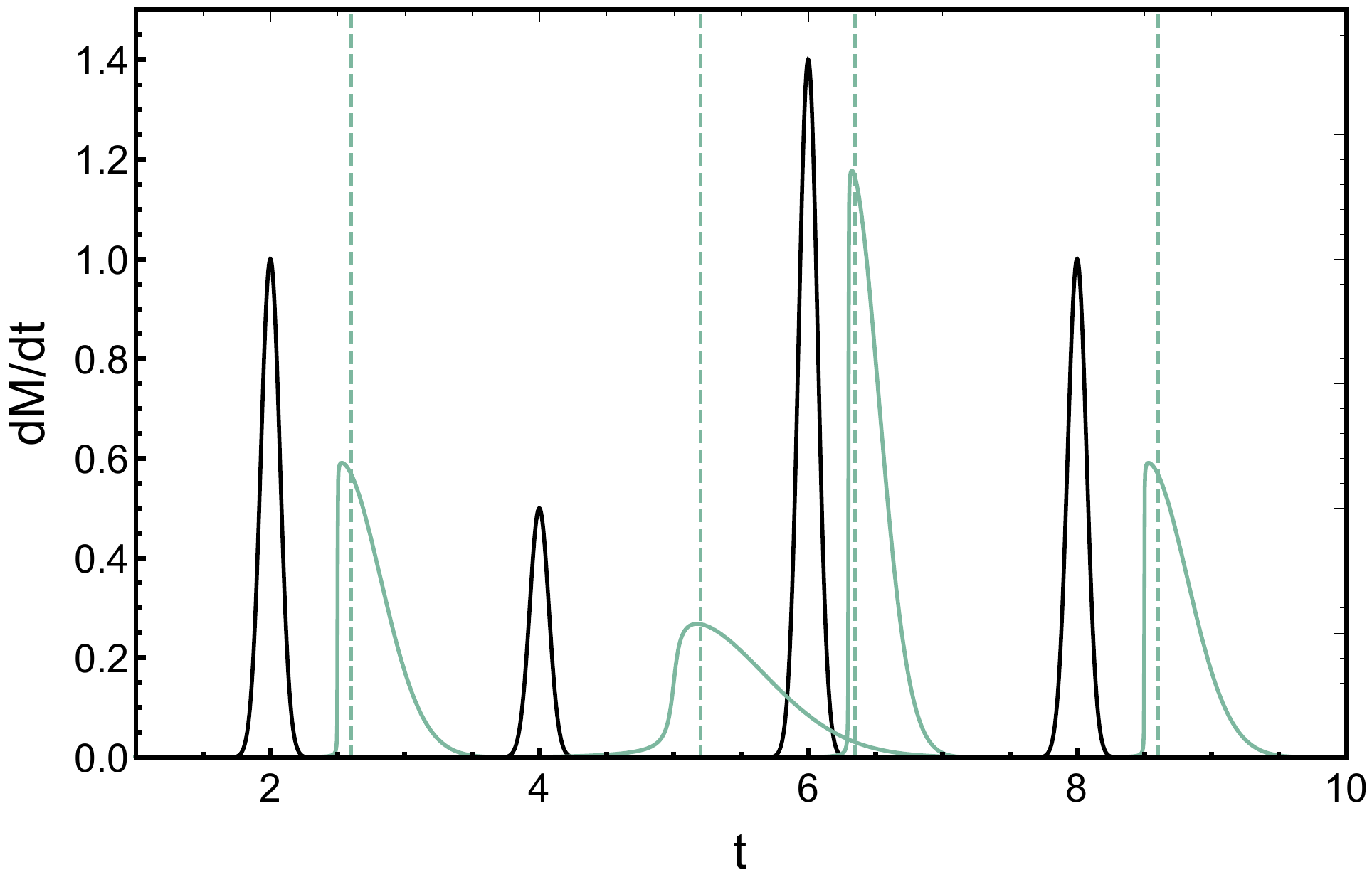}
    \caption{Even nearly periodic episodes of mass-loss due to EMRI flybys can give rise to aperiodic flares if the viscous timescale of the gaseous disk is variable (Eq.~\ref{eq:TQPEtot}).  This is illustrated schematically here, where we have introduced a variable delayed smoothing kernel (mimicking the viscous response of a gaseous disk) to produce an aperiodic signal ({\it green lines}) from a strictly periodic injection ({\it black lines}).}
    \label{fig:nonperiodic}
\end{figure}

The QPE source GSN 069 \citep{Miniutti+19} exhibited an $\approx 8\%$ increase in its period $T_{\rm QPE}$ over several months of observations, while RXJ 1301.9+2747 exhibited a $\approx 50\%$ change in the peak-to-peak interval between two consecutive bursts \citep{Giustini+20}.  While the interacting EMRI scenario so far described would predict slow changes in $T_{\rm QPE}$ due to evolution of the stellar orbits (from gravitational wave emission or angular momentum transfer with gaseous material), the timescale for significant changes $\sim t_{\rm dest}$ is much longer than these observed changes.  

However, one must consider possible variation in the delay between the release of gas by the (strictly) periodic tidal flyby, and the subsequent accretion onto the SMBH (see Fig.~\ref{fig:nonperiodic}).  The interval between flares is actually the sum of the flyby period and the viscous timescale of the gaseous disk, i.e.
\be
T_{\rm QPE} \simeq T_{\rm fly} + \tau_{\rm QPE}
\label{eq:TQPEtot}
\ee
The viscous timescale depends on the scale-height of the gaseous disk $\tau_{\rm QPE} \propto (h/r)^{-2}$ (Eq.~\ref{eq:tauQPE}), which in turn depends sensitively on the accretion rate.  Larger accretion rates lead to thicker disks (larger $h/r$) and hence shorter $\tau_{\rm QPE}$.  Such a scenario would predict shorter $T_{\rm QPE}$ following larger amplitude flares, consistent with the observed trend of increasing $T_{\rm QPE}$ and decreasing $L_{\rm QPE}$ in GSN 069 \citep{Miniutti+19}.

Several mechanisms could give rise to stochastic or secular evolution in the amount of mass-loss per flyby.  If either EMRI were to possess a small eccentricity, modulations in the stellar separation between flybys will lead to large variations in the tidally-stripped mass-loss (Appendix \ref{sec:flyby}).  Tidal forces by the companion could also excite periodic oscillations in the mass-losing star, rendering the amount of mass-loss sensitive to the phase (amplitude) of the oscillation at the time of the flyby.  The accretion timescale $\tau_{\rm QPE}$ (and hence $T_{\rm QPE}$) is also sensitive to the circularization radius of the gaseous debris, which depends on its (complex) interaction with the pre-existing gaseous disk. 

\subsection{Host Galaxy Nuclei}
\label{sec:galaxies}

While the nuclei of the {\it eROSITA} QPE hosts appear to be inactive \citep{Arcodia+21}, the first two QPEs occurred in galaxies with active nuclei possessing narrow emission-line regions \citep{Miniutti+19,Giustini+20}.  It is thus of interest to ask whether these galaxies are ``intrinsically'' active due to a pre-existing AGN disk, or whether the long-lived phase of accretion due to the interacting EMRIs studied here could power their activity.  

Accretion of the total stellar mass $M_2 \sim M_{\odot}$ over the active duration $t_{\rm dest} \sim 10^{3}-10^{4}$ yr (Eq.~\ref{eq:tdest}) will release a total energy $E \sim 0.1 M_2 c^{2} \approx 2\times 10^{53}m_2$ erg in UV/X-ray radiation, sufficient to ionize $M_{\rm ion} \sim (E/\epsilon_{\rm Ryd})m_p \sim 7\times 10^{6}m_2 M_{\odot}$ of hydrogen, where $\epsilon_{\rm Ryd} \simeq 13.6$ eV is the Rydberg and $m_p$ the proton mass.  This is broadly consistent with the inferred masses of Seyfert 2 narrow-line regions, while the radial extent of the predicted transient narrow-line region $\sim c t_{\rm dest} \sim 0.3-3$ kpc is also typical (e.g., \citealt{Vaona+12}).  It thus appears possible that a system of interacting EMRIs, if caught sufficiently late in their evolution, could generate its own transient narrow-line region.  However, the presence of a pre-existing gaseous AGN disk can help facilitate the migration of circular EMRIs into galactic nuclei (Section \ref{sec:AGN}) and hence the preferential occurrence of QPEs in intrinsic AGN environments might also be expected.

\subsection{Rate Estimates}
\label{sec:rates}

Here we provide rough estimates for the rate of circular EMRI formation needed to explain the observed QPE population.  Motivated by the blind nature of the {\it eROSITA} survey, we focus on the QPE discoveries in otherwise quiescent galactic nuclei.\footnote{QPE rates in AGN are harder to quantify without a treatment of selection effects that is beyond the scope of this paper.}  The first EMRI $M_1$ undergoes RLOF evolution over a timescale $\tau_{\rm GW} \sim 1-10$ Myr (Eq.~\ref{eq:tauGW}) for $M_1 \sim 0.3-1$.  This is longer than the interval between consecutive EMRIs if the latter occur at a per-galaxy rate $\dot{N}_{\rm EMRI} \gtrsim 1/\tau_{\rm GW} \sim 10^{-7}-10^{-6}$ yr$^{-1}$.  

The co-moving volume within the redshift $z = 0.0505$ of the most distant {\it eROSITA} source (eRO-QPE1) is $\mathcal{V} \approx 0.04$ Gpc$^{3}$.  Using the local density of Milky Way (MW)-like galaxies of $\mathcal{R}_{\star} \sim 6\times 10^{6}$ Gpc$^{-3}$ as a proxy for potential QPE hosts, and assuming a QPE active lifetime $\tau_{\rm dest}$ (Eq.~\ref{eq:tdest}), the number of QPEs in the survey can be estimated as,
\begin{eqnarray}
&N_{\rm QPE}& \sim \dot{N}_{\rm EMRI}f_{\rm cop} \mathcal{V}\tau_{\rm dest} \nonumber \\
&\sim& 2.4 \left(\frac{\dot{N}_{\rm EMRI}}{10^{-7}\,\rm yr^{-1}}\right) \left(\frac{f_{\rm cop}}{10^{-2}}\right)\left(\frac{\tau_{\rm dest}}{10^{4}{\rm \,yr}}\right),
\label{eq:NQPE}
\end{eqnarray}
where $f_{\rm cop}$ is the fraction of the EMRIs that are coplanar to within the range of mutual inclination $i \lesssim i_{\rm max} \sim 5R_{2}/a \sim 10^{-2}$ that permit interactions of the type required to generate strong periodic mass-loss.

The relevant value of $f_{\rm cop}$ depends on the EMRI formation channel.  While $f_{\rm cop} \sim i_{\rm max} \sim 10^{-2}$ for EMRIs that arrive with an isotropic distribution of inclination angles, we could expect $f_{\rm cop} \sim 1$ for EMRIs that arrive by migrating through a gaseous AGN disk (Section \ref{sec:AGN}).  

However, this simple rate estimate is complicated by general relativistic nodal precession of EMRI orbits inclined with respect to the SMBH spin (Section \ref{sec:precession}).  While precession will bring even initially misaligned EMRIs into temporary alignment (on the precession timescale $T_{\rm prec}$ of months to years; Eq.~\ref{eq:Tprec}), the limited duty cycle of the alignment compensates by increasing the required rate by a factor $\sim i_{\rm max}^{-1} \sim 10^{2}$, so one is back to $f_{\rm cop} \sim 10^{-2}$ as in the isotropic case.  On the other hand, if the observed QPE population exhibits no evidence for precession  (e.g., as a ``turn-off'' of the QPE signal on a timescale $\sim T_{\rm prec}$), then the required ``double coincidence'', namely that the EMRI orbits must be aligned both with each other as well as with the SMBH spin, acts to increase the required rate by a larger factor $i_{\rm max}^{-2} \sim 10^{4}$ and hence one has an effective value of $f_{\rm cop} \sim 10^{-4}$ entering Eq.~\ref{eq:NQPE}.

In summary, the number of QPEs $N_{\rm QPE} = 2$ detected thus far by {\it eROSITA} \citep{Arcodia+21} requires a rate $\dot{N}_{\rm EMRI} \gtrsim 10^{-7}$ yr$^{-1}$ (for $f_{\rm cop} = 10^{-2}$; precessing case) and $\gtrsim 10^{-5}$ yr$^{-1}$ (for $f_{\rm cop} = 10^{-4}$; non-precessing case).  The next section explores different circular EMRI channels and to what extent they can generate these rates.

\subsection{Conservative Mass-Transfer?}

Our calculations thus far have neglected changes in the orbital separation between $M_1$ and $M_2$ that arise if the angular momentum of the stripped debris during flybys is transferred back into one or the other orbit.  In the most extreme case in which 100\% of the angular momentum is transferred back to $M_2$, the increase in orbital separation between $M_1$ and $M_2$ due to a loss of mass $\Delta m_{\rm fly}$ is given by (MS17)
\be
\frac{\Delta a_{m}}{R_2} \approx 2\frac{a}{R_2}\left(\frac{\Delta m_{\rm fly}}{M_2}\right) \approx 4\times 10^{-5}\frac{M_{\bullet,6}^{1/3}}{m_2^{1/3}}\left(\frac{\Delta m_{\rm fly}}{10^{-7}M_2}\right),
\ee
For typical ejecta masses $\Delta m_{\rm fly} \sim 10^{-8}-10^{-6}M_{\odot}$ (Eq.~\ref{eq:deltamdest}), $\Delta a_{m}$ can greatly exceed the compensating decrease in the stellar orbital separation between flybys driven by gravitational wave emission, $\Delta a_{\rm GW}/R_{2} \sim 10^{-7}$ (Eq.~\ref{eq:deltab}).  

If $\Delta a_{m} \gg \Delta a_{\rm GW}$ then the mass-loss rate per flyby will be regulated to a value smaller than we have calculated by neglecting angular momentum added back to the orbit.  Indeed, the limit of fully conservative mass-transfer, the time-averaged accretion luminosity is regulated to a value $ \langle L \rangle \sim 10^{39}-10^{41}$ erg s$^{-1}$ (Eq.~\ref{eq:Lavg}) which, as already mentioned, is too small to explain observed QPE luminosities, $\langle L_{\rm QPE} \rangle \gtrsim 10^{42}$ erg s$^{-1}$.

However, there are several reasons why one would expect mass transfer onto the SMBH to be highly non-conservative in this environment.  For instance, the vertical scale-height of the gaseous disk $h/r \gtrsim 0.1$ (as necessary to explain the short durations of QPE flares; Section \ref{sec:timescale}) greatly exceeds $M_2$'s Hill radius $\sim R_2/a \sim 0.01$ (Eq.~\ref{eq:aRL}).  

Furthermore, even if the mass transfer is fully conservative and the accretion luminosity is fixed to the value $\langle L \rangle \ll \langle L_{\rm QPE} \rangle$ (Eq.~\ref{eq:Lavg}), limited periods of much higher mass-transfer are allowed as long as they are compensated by much longer periods at lower $\dot{M}$.  This very situation will occurs as a result of the EMRI orbital precession (Section \ref{sec:precession}), which results in an active duty cycle $\sim 5 \Delta R_2/a \sim 0.01$ for flyby-induced flares when the orbits are aligned in a common plane.  
Increasing $\langle L \rangle$ (Eq.~\ref{eq:Lavg}) by a factor $\sim 100$ during the comparatively brief active window results in accretion luminosities $\sim 10^{42}-10^{43}$ erg s$^{-1}$, consistent with the time-averaged QPE luminosities.  Thus, although the detailed predictions will change, whether the mass transfer is fully conservative or non-conservative, interacting EMRI systems (when active) can produce mass-transfer rates in broad agreement with QPE observations.

\section{Formation Channels}
\label{sec:channels}

The primary challenge for our QPE scenario, and for alternative single EMRI-related explanations (Section \ref{sec:alternative}), is to bring stars onto tightly bound orbits without destroying them through tidal disruption or energy deposition in the process. In this section, we discuss potential channels for generating consecutive co-planar EMRIs.  We ultimately find two possibilities to be the most promising: (i) relativistic circularization following the destruction of a binary through the Hills mechanism (Section \ref{sec:Hills}), and (ii) quasi-circular migration through an AGN disk (Section \ref{sec:AGN}).

\subsection{Dynamical Channels}
\label{sec:dynamical}

\subsubsection{General Constraints}
Stars can not be formed {\it in situ} on scales of $\sim R_{\rm RL} \sim 10^2 R_{\rm g}$; they must be delivered from larger radii.  If this occurs through high-eccentricity migration from some initial semimajor axis $a_0$, then GW emission is the only way the orbit can circularize.  Internal tidal dissipation would blow up the star long before circularization, as $GM_\bullet/R_{\rm RL} \gg GM_\bullet/a_0$.  

The time required to substantially circularize an orbit of initial eccentricity $e_0 \approx 1$ via GW emission is \citep{Peters64}
\begin{equation}
    T_{\rm GW}^{\rm ecc} \approx \frac{24\sqrt{2}}{85} \frac{a_0^{1/2}q_0^{7/2} c^5}{G^3 M_\bullet M_1 },
\end{equation}
where as before we consider a star with initial mass $M_1$ and radius $R_1$, and an initial pericenter $q_0 = a_0 (1-e_0)$.  If orbital perturbations do not affect the star's angular momentum on timescales $\gtrsim T_{\rm GW}^{\rm ecc}$, then the star can circularize and eventually become a RLOF EMRI.  However, rapid perturbations to the stellar orbital angular momentum will either move it to larger pericenter $q$ (aborting the circularization process) or smaller $q$ (resulting in the disruption of the star if $q < R_{\rm t}\equiv R_1(M_\bullet / M_1)^{1/3}$, the parabolic tidal disruption radius).  The most generic source of angular momentum perturbations is two-body non-resonant relaxation\footnote{At the small pericenters $q \lesssim 100 R_{\rm g}$ considered here, scalar resonant relaxation will likely be detuned by general relativistic precession.  Stronger secular torques from axisymmetric features of the nuclear potential, such as stellar disks, may persist, but are beyond the scope of this work.}, which operates on a star of eccentricity $e$ on the timescale 
\begin{align}
    T_{\rm AM} \approx& T_{\rm r} (1-e^2) \approx \frac{0.3}{\ln\Lambda}\frac{\sigma^3(r)(1-e^2)}{G^2 n(r) \langle m^2 \rangle} \notag \\
    \approx & \frac{0.75}{(3-\gamma)(1+\gamma)^{3/2}}\frac{M_\bullet^{1/2} \langle m\rangle r_{\rm infl}^{3-\gamma}}{G^{1/2} \langle m^2 \rangle \ln \Lambda}a^{\gamma-5/2}q. 
\end{align}
Here we have used the standard energy relaxation time $T_{\rm r}$ for an assumed power-law stellar density profile $n(r) \propto r^{-\gamma}$, which has a 1D velocity dispersion $\sigma(r) = \sqrt{GM_\bullet / r} / (1+\gamma)$.  We have defined: the influence radius $r_{\rm infl}$ inside of which the enclosed stellar mass equals $M_\bullet$; the first ($\langle m \rangle$) and second ($\langle m^2 \rangle$) moments of the stellar present-day mass function (PDMF); and the Coulomb logarithm $\ln \Lambda \approx \ln(0.4 M_\bullet / \langle m \rangle)$.

Requiring $T_{\rm AM} > T_{\rm GW}^{\rm ecc}$ for a stellar EMRI can be translated into the condition that $q_0 < q_{\rm GW}$, where
\begin{equation}
    q_{\rm GW} \approx 3.2 R_{\rm g} \xi^{-2/5} \zeta^{2/5} \left( \frac{a_0}{r_{\rm infl}} \right)^{-(6-2\gamma)/5},
\end{equation}
and we have defined $\xi \equiv (1+\gamma)^{3/2}(3-\gamma) \ln \Lambda$ and $\zeta \equiv M_1 \langle m \rangle / \langle m^2 \rangle$.  We note that $\xi \sim 10$ always, and $\zeta \sim 1$ usually, although $\zeta \ll 1$ for very low mass target stars and/or PDMFs rich in stellar mass black holes.  

For GW circularization to be possible, we also require that $q_{\rm GW} > R_{\rm t}$, which places an upper limit on the initial semimajor axis: $a_0 < a_{\rm GW}$, where
\begin{equation}
    \frac{a_{\rm GW}}{r_{\rm infl}} \approx (0.0028)^{\frac{1}{6-2\gamma}} \left(\frac{\zeta}{\xi}\right)^{\frac{1}{3-\gamma}} \left( \frac{R_{\rm g}}{R_1} \right)^{\frac{5}{6-2\gamma}} \left(\frac{M_\bullet }{M_1} \right)^{\frac{5}{6\gamma}-18}. \label{eq:a0}
\end{equation}
Assuming the constraints $a_0 < a_{\rm GW}$ and $q_0 < q_{\rm GW}$ indeed hold, we can now compute the residual eccentricity, $e_{\rm res}$, left over at the beginning of RLOF.  We do this by making use of the \citet{Peters64} constant of motion $c_0 = a(1-e^2)e^{-12/19}(1+121e^2/304)^{-870/2299}$, assuming that the initial $1-e_0 \ll 1$, and the final residual (i.e., beginning of RLOF) eccentricity $e_{\rm res} \ll 1$.  This yields
\begin{align}
    e_{\rm res} \approx& 0.22 \left( \frac{\xi}{\zeta} \right)^{19/30} \left( \frac{q_0}{q_{\rm GW}} \right)^{-19/12} \left( \frac{R_1}{R_{\rm g}} \right)^{19/12} \label{eq:residual} \\
    &\left( \frac{M_\bullet}{M_1} \right)^{19/36} \left(\frac{a_0}{r_{\rm infl}} \right)^{19(3-\gamma)/30}    .\notag
\end{align}
Setting $a_0 = a_{\rm GW}$, we thus obtain a simple expression for the maximum value of the residual eccentricity:
\begin{equation}
    e_{\rm res} \le e_{\rm res}^{\rm max} \approx 0.034 \left( \frac{q_0}{q_{\rm GW}} \right)^{-19/12}.  
    \label{eq:emax}
\end{equation}
Interestingly, this value is large enough to produce substantial variation of the QPE amplitude between eruptions (Eq.~\ref{eq:fecc}).  

\subsubsection{Single-Star Scattering}
\label{sec:singlestar}

How do stars find themselves on sufficiently tight orbits to satisfy the constraint $a_0 < a_{\rm GW}$ posed in the previous section?  In a power-law galactic nucleus, the flux of stars into the loss cone, from a bin of semimajor axis $a$, driven by two-body relaxation, is roughly $\mathcal{F}(a) \propto a^{9/2-2\gamma}$ \citep{Stone&Metzger16}.  Since the peak of loss cone flux is sourced from $a \sim r_{\rm infl}$, this implies that the stellar EMRI rate from two-body relaxation will be suppressed by a factor $\sim (a_0 / r_{\rm infl})^{9/2-2\gamma}$ relative to the total rate of tidal disruption events (TDE).  Taking a Bahcall-Wolf cusp with $\gamma=7/4$, and $a_0 / r_{\rm infl} \sim 10^{-4}$ (as is implied by Eq. \ref{eq:a0} for $\gamma=7/4$ and $M_\bullet = 10^6 M_\odot$), this rate suppression is $\sim 10^{-4}$.  Considering that the average per-galaxy rate of TDE is $\sim 10^{-4}~{\rm yr}^{-1}$ \citep{Stone&Metzger16, vanVelzen18}, the rate of single-star EMRIs with properties capable of generating QPEs is at least an order of magnitude too low to explain the {\it eROSITA} detections (Eq.~\ref{eq:NQPE}).

The single-scattering scenario also runs into the problem of whether it is even realistic for a Bahcall-Wolf cusp of stars to exist down to such small semi-major axes $\sim a_{\rm GW} \sim 10^{-4}{\rm pc} \ll r_{\rm inf}.$  In-situ star formation is unlikely at such small radii because no AGN disk would be gravitationally unstable so close to the SMBH.  Diffusion in energy space is also problematic because (non-compact) stars, for which their surface escape speed is less than the local dispersion velocity, are just as likely to undergo physical collisions as to be placed onto such tight orbits through 2-body scattering (e.g., \citealt{Frank&Rees76}).  The only robust mechanism for placing stars onto orbits with such small semi-major axes is the Hills mechanism, which is the focus of the next section.

\subsubsection{Hills Mechanism}
\label{sec:Hills}

The challenge of producing stellar EMRIs from two-body relaxation alone has motivated past work to consider the Hills mechanism \citep{Hills88}.  If a binary star with initial internal semimajor axis $A_{\rm bin}$, and total mass $M_{\rm bin}$, approaches the SMBH on a highly radial orbit, it will be tidally detached if its external pericenter $q_{\rm bin}$ is smaller than the binary detachment radius $R_{\rm bin} \approx A_{\rm bin}(M_\bullet / M_{\rm bin})^{1/3}$.  One binary component will be ejected as a hypervelocity star (see \citealt{Koposov+20} for a recent example from our own Galactic center), while the other star will become bound to the SMBH with pericenter $q_0 \sim q_{\rm bin}$ and a semimajor axis $a_0 \gtrsim a_{\rm Hills}$, where
\begin{equation}
    a_{\rm Hills} = \frac{A_{\rm bin}}{2} \left( \frac{M_\bullet}{M_{\rm bin}}\right)^{2/3}.
\end{equation}
In practice, $a_0 \sim a_{\rm Hills}$ usually, although in a minority of cases $a_0 \gg a_{\rm Hills}$ will occur.  

The Hills mechanism is an appealing way to produce tightly bound stars with $a_0 < a_{\rm GW}$, because the binaries can approach the SMBH from arbitrarily far away: the post-detachment $a_0$ of the bound star is determined primarily by $A_{\rm bin}$.  Indeed, if we require that $a_0 < a_{\rm GW}$, we find that 
\begin{equation}
    \frac{A_{\rm bin}}{r_{\rm infl}} < 2(0.0028)^{\frac{1}{6-2\gamma}} \left(\frac{\zeta}{\xi}\right)^{\frac{1}{3-\gamma}} \left( \frac{R_{\rm g}}{R_1} \right)^{\frac{5}{6-2\gamma}} \left(\frac{M_\bullet }{M_1} \right)^{-\frac{11-2\gamma}{18-6\gamma}},
\end{equation}
where we have approximated $M_{\rm bin} \approx M_1$.  For Bahcall-Wolf cusps ($\gamma = 7/4$) and main sequence stars, this requirement (for $\xi = 10$; $\zeta = 1$), \be
A_{\rm bin} \lesssim 0.6 R_{\odot} M_{\bullet,6}r_{1}^{2}m_1 \left(\frac{r_{\rm infl}}{0.1\,\rm pc}\right),
\label{eq:Abinmax}
\ee
becomes quite restrictive and limits us to considering extremely tight binaries with $A_{\rm bin} \lesssim R_\odot$.  

The rate of such ``Hills EMRIs'' is quite uncertain.  The total rate of binary separations in simple, spherically symmetric models for a MW-like galactic nucleus is $\sim 10^{-5}~{\rm yr^{-1}} (f_{\rm B} / 0.1)$, where $f_{\rm B}$ is the binary fraction \citep{YuTremaine03}.  This rate can increase by one to two orders of magnitude if the regions outside the SMBH influence radius have a strongly triaxial geometry \citep{MerrittPoon04}, or contain massive perturbers such as giant molecular clouds \citep{Perets+09} or nuclear spiral arms \citep{Hamers&Perets17}.  However, only $\sim 1\%$ of post-detachment stars will successfully evolve into a quasi-circular EMRI, as it is much more common for the post-detachment $a > a_{\rm GW}$ \citep{AmaroSeoane+12}.  

At the order of magnitude level, we thus expect the rate of quasi-circular EMRIs sourced from the total nuclear stellar population in MW-like galaxies to be $\sim 10^{-7} - 10^{-5}~{\rm yr}^{-1}$.  This is $1-3$ orders of magnitude higher than the single star scattering rate and broadly consistent that needed to explain the observed {\it eROSITA} QPEs (Section \ref{sec:rates}) for typical source lifetimes $\tau_{\rm dest} \sim 10^{4}$ yrs.  

An alternative route to producing the observed {\it eROSITA} QPE sample is to rely on secular dynamics in nuclear stellar disks.  The S stars in the center of the MW may be the bound byproducts of the Hills mechanism, operating on binaries originating in a sub-pc disk of stars (e.g., \citealt{Madigan+14,Generozov&Madigan20}).  Secular torques produced by global eccentricity features of the disk can quickly excite binaries to radial orbits that are vulnerable to the Hills mechanism.  The production of $\approx 100$ S stars in the last $\approx 10^7$ yr indicates a time-averaged binary detachment rate of $\sim 10^{-5}~{\rm yr}^{-1}$ in MW-like galaxies, similar to the minimum rates calculated above.  One appeal of the disk scenario, however, is that a significant fraction of the bound binary components may orbit the SMBH in roughly the same orbital plane.  In particular, secular eccentricity excitations of disk members are accompanied by inclination excitation, which \citet{Wernke&Madigan19} find results in $\sim 10-20\%$ of the orbits at disruption being inclined in a narrow range of angles centered around $\pm 180^{\circ}$ relative to the stellar disk.%

\subsection{AGN Migration}
\label{sec:AGN}

\begin{figure}
    \centering
    \includegraphics[width=0.48\textwidth]{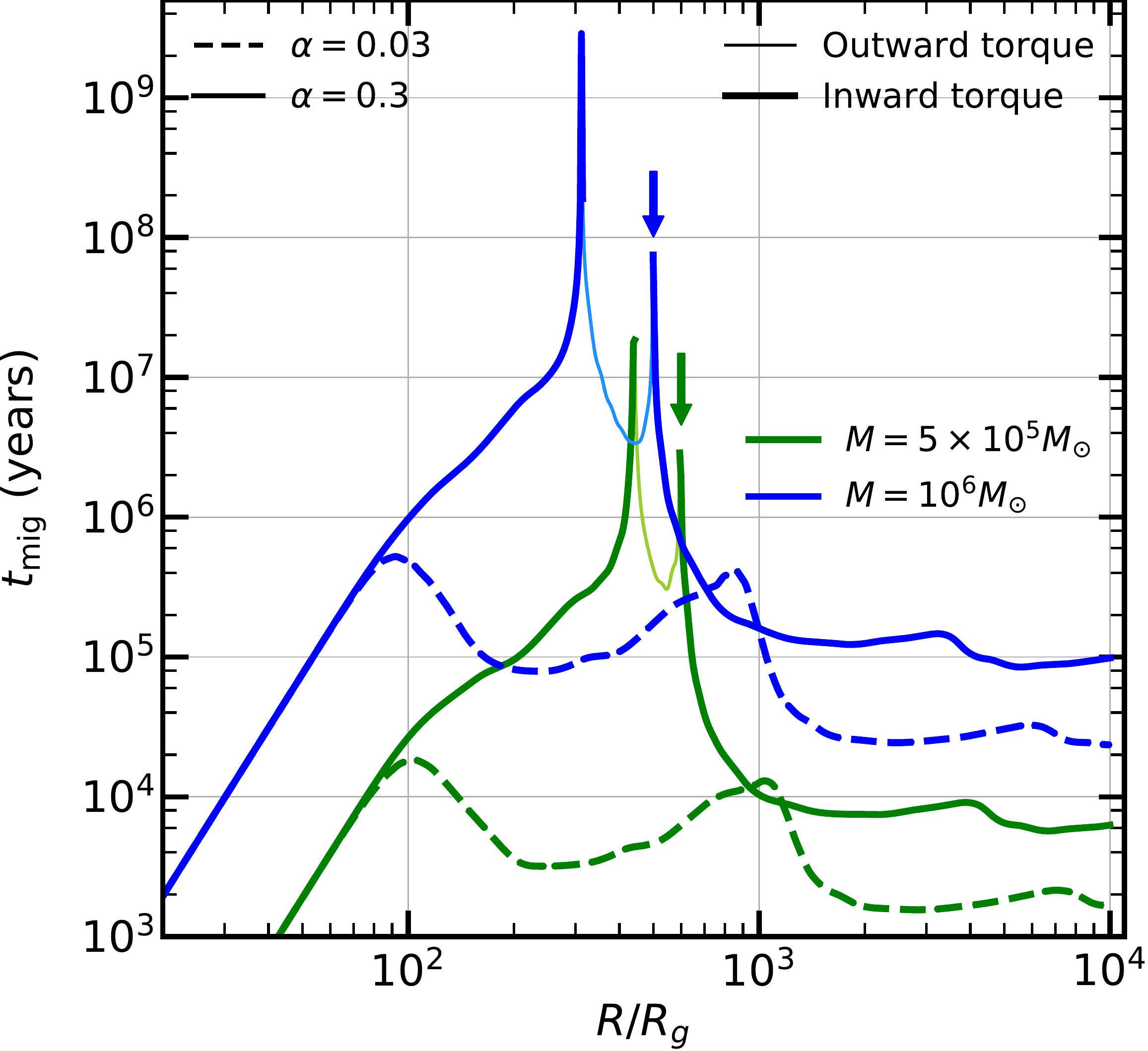}
    \caption{Migration time scale ($R/\frac{da}{dt}$), due to the combination of GW emission and Type I gaseous torques, as a function of radius in a \citet{ShakuraSunyaev73} AGN disk calculated for a $1M_\odot$ migrating star.  We show for two limiting cases of the effective viscosity parameter $\alpha =0.03$ (dashed) and $\alpha=0.3$ (solid), and two SMBH masses: $10^6M_\odot$ (blue) and $5\times 10^5 M_\odot$ (green). We see that both SMBH masses result in migration traps (indicated by arrows) for $\alpha=0.3$ where the inward migration(thick line) meets the outward migration (off colour thin line).}
    \label{fig:migration}
\end{figure}
Another source for producing circular EMRIs is via inwards migration of stars embedded within a gaseous AGN disk \citep{Levin07}.  Stars may form {\it in situ} in AGN disks \citep{SirkoGoodman03} or alternatively may be captured by gas drag \citep{Syer+91}.  Regardless of their origin, once embedded within the disk (typically at radii $\gg 1$ AU), they will migrate inwards, primarily due to ``Type I'' torques \citep{GoldreichTremaine79}.

While Type I migration is generally inwards, the gaseous torque can flip sign in small regions of AGN disks, in a way that depends on the accretion rate, viscosity, and radial location within the disk \citep{Paardekooper+10}.  In particular, radial zones can develop where the net migration torque is positive, causing outwards migration.  The interface between an inner region exerting positive torque and an exterior region exerting a negative torque is a ``migration traps'' where many stellar mass objects can accumulate \citep{Bellovary+16}.  We estimate the location of these traps in low-mass AGN using a \citet{ShakuraSunyaev73} disk model with realistic opacities \citep{Rogers+96} and computing Type I torques following \citet{Paardekooper+10}.  As shown in Figure \ref{fig:migration}, migration traps naturally develop on scales $\sim 10^{2-3} R_{\rm g}$ for SMBH masses $M_{\bullet} \sim 3\times 10^{5}-3\times 10^{6}M_{\odot}$ characteristic of QPE host galaxies. 

In active galaxies, migration traps on $\sim$ AU (i.e. $\sim 10^2 R_{\rm g}$) scales offer a natural mechanism for producing the coplanar EMRI pairs needed to generate QPEs.  The simplest version of this mechanism involves a star (or compact object) parked in the migration trap, following a Type I inspiral that is prograde with respect to the gas.  The subsequent retrograde inspiral\footnote{For a star orbiting retrograde with respect to the AGN disk, the dominant torque will not be Type I migration (which does not exist for retrograde orbiters) but closer to gas dynamical friction.  While the second object will not be caught in the trap itself, the gas dynamical friction time is likely $\gg t_{\rm dest}$.} of an outer star eventually triggers its RLOF, with the QPEs beginning once the two stars approach sufficiently closely.  One uncertainty in this formation channel concerns the modifications to RLOF for stars embedded in AGN gas.  For stars that are on prograde orbits, these modifications are likely modest, as the subsonic AGN gas will simply add an extra pressure force to the outer boundary condition of the star.  For stars on retrograde orbits, however, the supersonic ram pressure of the AGN gas may play a role in ablating the stellar atmosphere (Appendix \ref{sec:ablation}).  

Even in the absence of retrograde orbiters, migration traps can accumulate large chains of prograde stars trapped in mean motion resonances \citep{Secunda+19}.  After the AGN episode ends, this chain of stars could begin to undergo GW migration inward, setting off a sequence of co-orbiting EMRIs (Section \ref{sec:ASASSN}).  This could result in a preference for QPEs or other periodic nuclear sources in post-starburst galaxies that have undergone major mergers and associated AGN activity in the relatively recent past (indeed, RX J1301.9+2747 is hosted by a post-starburst galaxy; \citealt{Giustini+20}).

\section{Discussion}
\label{sec:discussion}

\subsection{Long Duration QPEs from Co-orbiting EMRIs}
\label{sec:ASASSN}

Counter-orbiting EMRIs are favored as the origin of the recently discovered X-ray QPEs due to the need to generate intervals between close flybys of less than a day (see Eq.~\ref{eq:taucoll1}).  However, co-orbiting collisions should also occur, which for the same stellar parameters ($\rho \sim 0.1-10$ g cm$^{-3}$) would predict longer QPE periods of $T_{\rm QPE} \sim$ days to months and commensurately larger average ejecta masses $\Delta m_{\rm dest} \propto T_{\rm QPE}$ (Eq.~\ref{eq:deltamdest}).  

A candidate for such a long-period QPE is the periodically flaring AGN, ASASSN-14ko, which exhibits outbursts at regular intervals of around 114 days \citep{Payne+21}.   During the rise of its ouburst in May 2020, ASASSN-14ko exhibited UV bright, thermal spectral energy distribution similar to tidal disruption events.  However, the X-ray flux decreased by a factor of $\approx$ 4 at the beginning of the outburst before returning to its quiescent flux after $\sim$8 days.  The large inferred black hole mass $M_{\bullet} \sim 10^{8}M_{\odot}$ for ASASSN-14ko would require two stars with $\rho \lesssim 0.1 \rho_{\odot}$ for RLOF to occur outside the ISCO radius.  For the same parameters, Eq.~(\ref{eq:taucoll2}) predicts a QPE period of $T_{\rm QPE} \sim 100$ d, consistent with the activity period in ASASSN-14ko.  The EMRI destruction time for such a massive SMBH can be relatively relatively short, $t_{\rm destr} \sim 100$ yr (Eq.~\ref{eq:tdest}), and so appreciable evolution of the system could be observable.

\subsection{RLOF from single eccentric EMRI?}
\label{sec:alternative}

\citet{Zalamea+10} consider a scenario in which a WD on an eccentric orbit undergoes periodic RLOF (or, equivalently, partial tidal disruption) onto a SMBH, feeding gas onto the SMBH and powering a quasi-periodic string of flares.  A variant of this scenario was proposed in \citet{King20}, and both could, in principle, appear as QPEs \citep{Arcodia+21}.  This eccentric single-star partial disruption scenario requires a high-eccentricity orbit for a typical WD density $\rho \sim 3\times 10^{5}\rho_{\odot}$, as the orbital period at this object's Roche radius is only $\sim 1$ minute (Eq.~\ref{eq:tauorb}).  Reproducing observed QPE periods of $\approx 2- 19$ hours would thus require orbits with eccentricity $e \gtrsim 0.9-0.99$.  

One challenge to this scenario is the difficulty of putting a single star on a significantly eccentric orbit with a semimajor axis of $\sim 1$ AU.  \citet{King20} invokes the tidal disruption of the envelope of a red giant star, which leaves behind a degenerate core on an orbit with appropriate parameters.  This version of the single-star scenario has two major challenges:
\begin{enumerate}
    \item The magnitude of the ``core kick'' is far too low to put a surviving core on an orbit with period $\sim T_{\rm QPE}$.  The origin of core kicks lies in the deviation tensor (the third-order expansion of the gravitational potential), which encodes the asymmetries of the tidal field \citep{BrassartLuminet08, ChengEvans13}.  As this is simply the next-order expansion of the gravitational potential, beyond the second-order tidal expansion, we can estimate the magnitude of the specific orbital energy perturbation to the surviving core on dimensional grounds (in analogy to the reasoning of \citealt{Stone+13}) as
    \begin{eqnarray}
        \delta \epsilon \sim \frac{GM_\bullet}{R_{\rm t}} \left(\frac{R_\star}{R_{\rm t}} \right)^2 \frac{\Delta M}{M_{\rm core}} 
        \sim \frac{GM_\star}{R_\star} \frac{\Delta M}{M_{\rm core}}.
        \label{Eq:deltakick}
    \end{eqnarray}
    Here, we have considered the partial disruption of a star with initial mass $M_\star$ and radius $R_\star$, which leaves behind a surviving core of mass $M_{\rm core} = M_\star - \Delta M$ following a close encounter near or inside the tidal radius $R_{\rm t} = R_\star (M_\bullet / M_\star)^{1/3}$.  This order of magnitude estimate is in good agreement with numerical hydrodynamic simulations of core kicks from partial disruptions \citep{Manukian+13, Gafton+15}.  For the disruption of a red giant ($M_{\star} \sim 1 M_{\odot}$; $R_{\star} \sim 200R_{\odot}$) with initial specific orbital energy $\epsilon_{\rm RG}$, it predicts a final specific energy for the bound core of $\epsilon_{\rm RG} \pm \delta \epsilon$ where $\delta \epsilon \sim 10^{-8} c^2$.  Since $\delta \epsilon \ll GM_\bullet / R_{\rm RL} \sim 0.01 c^2$, there is no way for the leftover core to have an orbital period comparable to $T_{\rm QPE}$.
    \item The sign of the core kick is likely positive-definite, i.e. $\delta \epsilon > 0$, so that surviving cores always become more loosely bound (rather than more tightly bound).  In the partial disruption simulations of \citet{Faber+05, Manukian+13, Gafton+15}, significant mass loss in a partial disruption is always associated with a positive energy kick to the surviving core.
\end{enumerate}

The production mechanisms and rates of eccentric WD EMRIs are not considered extensively by \citet{Zalamea+10}, although they suggest two-body scattering of a single WD or the Hills mechanism separating a binary with at least one WD component.  Both of these possibilities are disfavored on rates grounds.  For single-star scattering, we have already seen (\S\ref{sec:singlestar}) that rates are negligibly small, and they become even smaller when considering (relatively uncommon) WDs.  

This leaves the Hills mechanism as the favored way to produce stars with $a = (2\pi)^{-2/3} (G M_\bullet)^{1/3} T_{\rm QPE}^{2/3}$.  Recalling that the post-separation semi-major axis of the bound star $a_0 \gtrsim a_{\rm Hills}$, in order for a single star to be born into an orbit with a period equal to $T_{\rm QPE}$, the semi-major axis of the original binary must obey
\begin{equation}
    A_{\rm bin} \lesssim 0.1R_{\odot} \left( \frac{T_{\rm QPE}}{10~{\rm hr}} \right)^{2/3} \left( \frac{M_\bullet}{10^5 M_\odot} \right)^{-1/3} \left( \frac{M_{\rm bin}}{1.2 M_\odot} \right)^{2/3}. \label{eq:HillsInequality}
\end{equation}
This rules out the tidal detachment of a main sequence binary as capable of generating QPEs as short as 10 hours.  A binary composed of two WDs could satisfy Eq.~(\ref{eq:HillsInequality}), but any such tight binary has its own problem: a short lifetime.  The GW inspiral time of an equal-mass WD binary is only
\begin{equation}
    T_{\rm GW} \approx 3 \times 10^4~{\rm yr} ~ \left( \frac{A_{\rm bin}}{0.1R_{\odot}} \right)^4 \left( \frac{M_{\rm bin}}{1.2M_\odot} \right)^{-3},
    \label{eq:TGWbinary}
\end{equation}
rendering such systems exceedingly rare.  

Quantitatively, the rate of Hills separation of double WD binaries can be written, $\dot{N}_{\rm Hills}^{\rm WD} \sim \dot{N}_{\rm Hills} f_{\rm WD} f_{\rm hard}$, where $\dot{N}_{\rm Hills} \sim 10^{-5} - 10^{-3}~{\rm yr}^{-1}$ in a MW-type galaxy (Section \ref{sec:Hills}).  Here, $f_{\rm WD}$ is the fraction of all tidally detached binaries comprised of two WDs, and is likely $\lesssim 0.002$ (the total WD number fraction for a Salpeter IMF and an old stellar population).  The factor $f_{\rm hard}$ is the fraction of all double WD binaries with $A_{\rm bin}$ less than the critical value given by Eq.~(\ref{eq:HillsInequality}); lifetime arguments imply that $f_{\rm hard} < T_{\rm GW}/T_{\rm H} \sim 4\times 10^{-6}$, where $T_{\rm H}$ is the Hubble time.  Taken together, the total rate at which the Hills mechanism deposits single WDs onto sufficiently short-period orbits to explain the observed QPEs is $\dot{N}_{\rm Hills}^{\rm WD} \lesssim 10^{-10}~{\rm yr}^{-1}$ per MW-type galaxy, orders of magnitude below what is required by observations.

The rates problem is further exacerbated by the short predicted WD lifetime once Roche overflow starts.  Mass transfer onto the SMBH is likely to be unstable due to the inverted mass-radius relation of WDs and the non-conservative nature of mass transfer in highly eccentric binaries.  In the fiducial example given by \citet[their Fig.~2]{Zalamea+10}, the WD only loses a fraction $\sim 10^{-6}-10^{-8}$ of its mass (as required to explain QPE amplitudes) for a few hundred orbits.  This short lifetime $\sim 1$ month is in tension with archival X-ray detections of RX J1301.9+2747 and GSN 0691 going back decades \citep{Miniutti+19,Giustini+20}.  If the QPE lifetime is $\tau_{\rm dest} \lesssim 1$ yr, then the required WD EMRI rate to explain the {\it eROSITA} QPE sample is $\dot{N}_{\rm EMRI} \sim 10^{-5}$ gal yr$^{-1}$ (Eq.~\ref{eq:NQPE}), five orders of magnitude larger than we have estimated above for the Hills mechanism.

\section{Conclusions}
\label{sec:conclusions}

Building on previous work \citep{Metzger&Stone17}, we have proposed a mechanism for generating quasi-periodic eruptions in both active and otherwise inactive galactic nuclei through close flybys of stars on circular coplanar orbits, at least one of which is overflowing its Roche lobe onto the SMBH.  The latter requirement tightly constrains the model because, in the case of counter-orbiting stellar orbits, the observed QPE period is connected directly to the stellar structure (Fig.~\ref{fig:density}).  Although a large degree of uncertainty (both observational and theoretical) remains, our model naturally accommodates the range of observed QPE properties, including their periods (Eqs.~\ref{eq:taucoll1},\ref{eq:taucoll2}), durations (Eqs.~\ref{eq:tauQPE}), flare amplitudes (Eq.~\ref{eq:deltamdest}), QPE activity phase (driven by spin-induced orbital precession; Eq.~\ref{eq:Tprec}), total QPE active lifetimes (Eq.~\ref{eq:tdest}), and rates (Eq.~\ref{eq:NQPE}; Section \ref{sec:channels}).  Given the possible channels for generating circular EMRIs, we could expect the QPE phenomena in both inactive and active galactic nuclei, including those that are otherwise currently inactive but which formed stars through Toomre instability of an AGN disk in the relatively recent past.

One of the most stringent constraints on our model arises from the ``fragility'' of the RLOFing stars, which greatly limits the degree to which their orbits can evolve over the relatively short observational baselines of present QPE studies.  Although the orbital properties of the stellar pairs should remain almost strictly periodic over timescales of months to years, the QPE period itself may exhibit stochastic or possibly systematic changes, due to the additional hydrodynamic delay between the stellar mass loss and the accretion of gas by the SMBH (Fig.~\ref{fig:nonperiodic}).  Another prediction of our model is long-term modulation of the QPE signal due to SMBH spin-induced nodal precession, on a timescale of months to several years (Eq.~\ref{eq:Tprec}).  Interestingly, this ``turn-on'' and ``turn-off'  period of the QPE activity, if measured, could be used to constrain the spin of the central SMBH. 

Our model predicts gaseous disk masses (Eq.~\ref{eq:deltamdest}) that result in moderately sub-Eddington luminosity flares when accreted over QPE timescales.  However, we would expect order-of-magnitude variations in the peak accretion rate in different EMRI systems, extending to super-Eddington values.  If super-Eddington accretion generates relativistic jets, then we would predict (geometrically beamed) periodic hard X-ray flares, perhaps akin to longer lived, less luminous versions of jetted tidal disruption candidates such as Swift J1644+57 (e.g., \citealt{Bloom+11,Burrows+11}).  Quasi periodic non-thermal emission could in principle also be produced by plasmoids released into the accretion funnel by the passage of one EMRI through the misaligned gaseous disk (e.g., \citealt{Sukova+21}) generated by RLOF of the other EMRI (at epochs when the two EMRI orbital planes are not aligned) or by either EMRI passing through a pre-existing AGN. 

The circular EMRI systems we have described could also be detected through their periodic low frequency gravitational wave emission by space-based interferometers such as LISA.  Gravitational wave emission from ordinary (non-degenerate) stars undergoing RLOF is only detectable orbiting the SMBH in our own Galactic center \citep{Linial&Sari17}.  However, higher frequency emission from WD EMRIs provide a more promising extragalactic target \citep{Zalamea+10}, detectable by LISA out to several hundred Mpc distances (e.g., \citealt{Sesana+08}).  If QPE flares exist from WDs, then their higher mean densities result (in the counter-orbiting case) in periods $T_{\rm QPE} \sim 0.01-0.3$ hr (Fig.~\ref{fig:density}).  QPE with longer periods similar to those presently observed can be generated by {\it co-orbiting} WD EMRIs (Eq.~\ref{eq:taucoll2}); however, in this case, the gravitational wave frequency will greatly exceed the QPE frequency.  

Other scenarios involving single stars or WDs on eccentric orbits (\citealt{Zalamea+10,King20}), although nominally ``simpler'' than a two-EMRI model, run into serious difficulties explaining the QPE population.  Firstly, there is the general challenge of creating highly eccentric EMRIs without tidally destroying the star via tidal heating (Section \ref{sec:dynamical}).  Essentially impossible in the single-scattering dynamical channel (Eq.~\ref{eq:emax}), also in binary (Hills) scenarios one is limited to binaries with small semi-major axes (Eq.~\ref{eq:Abinmax}) such as WD binaries.  However, such tight WD EMRIs are short-lived due to their rapid GW inspiral times (Eq.~\ref{eq:TGWbinary}).  WD EMRIs are also unlikely to be produced by partial TDEs of giant stars due to the low expected kick on the bound core (Eq.~\ref{Eq:deltakick}).  Single EMRI scenarios, in which mass transfer is driven exclusively by GW radiation, cannot produce high enough mass-loss rates to explain the observed QPEs without being in a state of unstable (runaway) mass transfer, the short lifetimes of which further exacerbate the rate discrepancy.  

In addition to the $\sim$hours-day period QPEs generated by counter-orbiting EMRIs, our scenario predicts the existence of longer-period QPE-like periodic AGN from co-orbiting co-planar interacting EMRI pairs (Section \ref{sec:ASASSN}) or those with (non-precessing) misaligned orbital planes (MS17).

\acknowledgements

We thank the anonymous reviewer for their helpful insight and suggestions.  We acknowledge helpful conversations with Riccardo Arcodia, Aleksey Generozov, Zoltan Haiman, Yuri Levin, Ann-Marie Madigan, Itai Linial, Eliot Quataert, Mathieu Renzo, and Marta Volonteri.  This research was supported by through a NSF-BSF joint funding research grant (NSF grant AST-2009255 to BDM and BSF grant 2019772 to NCS and SG).  BDM acknowledges additional support from NASA (grant NNX17AK43G).  NCS acknowledges additional support from the Israel Science Foundation (Individual Research Grant 2565/19).

\appendix

\section{Enhanced Mass-Loss During EMRI Close Passages}
\label{sec:flyby}

Here, we estimate the influence of the gravity of $M_1$ on the mass-loss rate of $M_2$ during their flyby (when their radial separation $\Delta a = a_2 - a_1$), assuming $M_2$ is undergoing RLOF onto the SMBH, and that both orbits are circular.  First, we calculate the gravitational influence of $M_1$ in reducing the Hill radius of $M_2$.  Then we calculate the mass-loss rate from the brief-lived phase of enhanced RLOF.

Define a dimensionless Hill radius $x \equiv r_{\rm H}/r_{\rm H,0}$, where 
\be
r_{\rm H,0} \equiv \left(\frac{1}{3}\frac{M_2}{M_{\bullet}}\right)^{1/3}a_2
\ee
is the usual Hill radius, neglecting the effect of $M_1$ (Eq.~\ref{eq:aRL}).  Now, consider the influence of $M_1$ with a semi-major axis $a_1 < a_2$ and a temporary separation $\Delta a \equiv a_2 - a_1$ away from $M_2$ (which orbits with semi-major axis $a_2$).  The gravitational pull of $M_1$ will act to reduce the effective Hills radius to a value $r_{\rm H} \lesssim r_{\rm H, 0}$ (i.e. $x \lesssim 1$), although the deformation to the Hill sphere is asymmetrical, and the Hill sphere can actually grow along some angles.  Here we will consider the balance of gravitational
and centrifugal forces acting at distance $r_{\rm H}$ from $M_2$ along the common line connecting $M_1-M_2-M_{\bullet}$ at closest approach, can be written as
\be
\frac{GM_{2}}{r_{\rm H}^{2}}  -\frac{GM_{\bullet}}{(a_{2}-r_{\rm H})^{2}} - \frac{GM_{1}}{(\Delta a-r_{\rm H})^{2}} + \Omega^{2}(a_{2}-r_{\rm H}) = 0,
\ee
where the final term is the centrifugal force and $\Omega \simeq (GM_{\bullet}/a_{2}^{3})^{1/2}$.  Expanding this in the limits $M_1, M_2 \ll M_{\bullet}$, $\Delta a, r_{\rm H} \ll a_1, a_2$, we find
\be
\frac{M_{2}}{r_{\rm H}^{2}} - 3 \frac{M_{\bullet}}{a_{2}^{3}}r_{\rm H} - \frac{M_1}{(\Delta a -r_{\rm H})^{2}} = 0
\Rightarrow 
\frac{1}{x^{2}} - x - \frac{M_1}{M_2}\left(\frac{\Delta a}{r_{\rm H,0}} - x \right)^{-2} = 0
\ee
Defining $x \equiv 1-\epsilon$ with $\Delta a/r_{\rm H,0} \gg 1$ and $\epsilon \ll 1$, we have
\be
\epsilon \simeq \frac{M_1}{3M_2}\left(\frac{\Delta a}{r_{\rm H,0}}\right)^{-2} \simeq \frac{M_1}{3M_2}\left(\frac{\Delta a}{R_2}\right)^{-2}
\label{eq:epsilon}
\ee

Insofar as $M_2$ is filling its Roche lobe ($R_2 \simeq r_{\rm H,0}$), the close passage of $M_1$ causes the Roche surface to penetrate below the surface of $M_2$ around the $L_1$ point by a factor $\Delta r \simeq \epsilon R_2$.  Significant mass loss from $M_2$ through $L_1$, compared to the nominal mass-transfer rate onto the SMBH, will occur if $\Delta r \gg H$, where
\be
\frac{H}{R_2} = \frac{k R_2 T_{\rm eff}}{G M_2\mu m_p} \approx 7\times 10^{-4}\left(\frac{T_{\rm eff}}{10^{4}{\,\rm K}}\right)\frac{r_2}{m_2}
\label{eq:scaleheight}
\ee
is the density scale-height near the photosphere of $M_2$, $T_{\rm eff}$ is the stellar effective temperature and $\mu \approx 0.62$ the mean molecular weight.  We have normalized $T_{\rm eff}$ to a value $\sim 10^{4}$ K comparable to the expected value set by irradiation from the accretion flow,
\be T_{\rm eff} \approx \left(\frac{L_{\rm X}f_{\rm X}}{4\pi \sigma a^{2}}\right)^{1/4} \approx 1.6\times 10^{4}\,{\rm K}\,\left(\frac{L_{\rm X}}{10^{42}{\rm erg\,s^{-1}}}\right)^{1/4}\left(\frac{f_{\rm X}}{0.01}\right)^{1/4}\left(\frac{a}{1{\rm AU}}\right)^{-1/2},
\ee 
on a typical radial scales $a \sim 1$ AU (Eq.~\ref{eq:aRL}), where $L_{\rm X} \sim 10^{41}-10^{42}$ erg s$^{-1}$ is the time-averaged X-ray luminosities of QPEs (e.g., \citealt{Arcodia+21}) and the factor $f_{\rm X} \ll 1$ accounts for the (small) fraction of the total disk luminosity which reaches the orbital plane where $M_2$ resides. 

In general, the mass flow rate through the $L_1$ nozzle can be written as $\dot{m} \sim \rho c_{\rm s}r \Delta r$, where $\rho$ and $c_{\rm s}$ are the density and sound speed at depth $\Delta r$ inside the stellar atmosphere (e.g., \citealt{Lubow&Shu76}) and $r \Delta r$ is the nozzle's cross section at a distance $r$ from the center of the star.  The enhanced mass-loss rate during the closest passage of $M_1$ can then be expressed as (\citealt{Ritter88,Ginzburg&Quataert21})
\be
\dot{m}_{\rm fly} \sim \dot{m}_{\rm ph}\left(\frac{\Delta r}{H}\right)^{n+3/2},
\label{eq:mdotclose}
\ee
where $n$ is the effective polytropic index of the outer layers of $M_2$ ($n = 3/2$ for a convective region and $n = 3$ for a radiative region),
\begin{eqnarray}
\dot{m}_{\rm ph} \equiv \left(\frac{kT_{\rm eff}}{\mu m_p}\right)^{3/2}\frac{R_2^{3}}{GM_2}\rho_{\rm ph} \sim 3\times 10^{-16}{\rm M_{\odot} s^{-1}} \left(\frac{T_{\rm eff}}{10^{4}{\rm K}}\right)^{3/2}\left(\frac{\rho_{\rm ph}}{10^{-7}\rho}\right) \sim 1\times 10^{-16} {\rm M_{\odot} s^{-1}}  \frac{r_2}{\tilde{\kappa}}\left(\frac{T_{\rm eff}}{10^{4}{\rm K}}\right)^{1/2},
\label{eq:Mdotph}
\end{eqnarray}
and $\rho_{\rm ph}$ is the stellar photosphere density, which in the second equality is normalized to the mean stellar density $\rho = 3M_{2}/(4\pi R_{2}^{3})$ ($\rho_{\rm ph} \sim 10^{-7}-10^{-6}\rho$, typically).  In the final line of Eq.~(\ref{eq:Mdotph}) we have estimated the photosphere density as $\rho_{\rm ph} \simeq 1/H\kappa$, where $\kappa = \tilde{\kappa}\kappa_{\rm es}$ is the opacity normalized to that of electron scattering.
($\kappa_{\rm es} \simeq 0.38$ cm$^{2}$ g$^{-1}$), which gives
\be
\frac{\rho_{\rm ph}}{\rho}  \sim \frac{1}{\kappa H \rho} \simeq \frac{4\pi}{3}\frac{G\mu m_p R_{2}}{k T_{\rm eff}\kappa} \sim 4\times 10^{-8}\frac{r_2}{\tilde{\kappa}}\left(\frac{T_{\rm eff}}{10^{4}{\rm K}}\right)^{-1}
\ee

In both co-orbiting and counter-orbiting cases, the time interval $\tau_{\rm fly}$ over which the passing EMRIs spend near their closest approach (at stellar separation $\lesssim \Delta a$) can be written,
\be
\tau_{\rm fly} \sim \frac{\Delta a}{a}T_{\rm QPE},
\ee
where $T_{\rm QPE} = T_{\rm fly}$ (Eqs.~\ref{eq:taucoll1},\ref{eq:taucoll2}) is the time between flybys.

Combining results, the mass loss per flyby is given by
\begin{eqnarray}
\Delta m_{\rm fly} &\approx& \dot{m}_{\rm fly} \tau_{\rm fly} \approx \dot{m}_{\rm ph} \left(\frac{\epsilon}{H/R_2}\right)^{4.5}\tau_{\rm fly}
\approx 5\times 10^{-8}M_{\odot}\left(\frac{T_{\rm QPE}}{10\,{\rm hr}}\right)\left(\frac{T_{\rm eff}}{10^{4}{\rm K}}\right)^{-4}\frac{1}{\tilde{\kappa}}\frac{m_1^{4.5}}{r_2^{3.5}}\frac{m_2^{1/3}}{M_{\bullet,6}^{1/3}}\left(\frac{\Delta a}{5R_2}\right)^{-8},
\label{eq:deltam_app}
\end{eqnarray}
where we have assumed an $n = 3$ polytrope for the envelope structure of $M_2$, as expected due to the strong influence of irradiation from the SMBH accretion flow.  

So far we have assumed quasi-circular orbits, but we note here that large variations in $\Delta m_{\rm fly}$ will occur in the presence of relatively small residual eccentricities.  To generalize to the slightly eccentric case, we consider the inner star on a circular orbit, and the outer star on an orbit with eccentricity $e_2 \ll 1$.  The instantaneous separation at closest approach will be $\Delta r = \Delta a + a_2 e_2 \cos \psi$, where $\psi$ is a phase angle that varies stochastically from encounter to encounter\footnote{Correlated behavior of $\psi$ will only occur if the two stars are in mean motion resonance.}.  The mass loss in the flyby will now be the same as before, except $\Delta m_{\rm fly} \propto (\Delta r / R_2)^{-8}$.  We may therefore write a minimum mass loss, $\Delta m_{\rm fly}^{\rm min} \propto (\Delta a / R_2 + a_2 e_2/R_2)^{-8}$, and a maximum mass loss, $\Delta m_{\rm fly}^{\rm max} \propto (\Delta a / R_2 - a_2 e_2/R_2)^{-8}$.  The fractional difference between the circular -orbit limit and the maximum mass loss in an eccentric orbit will be \begin{equation}
    f_{\rm ecc} = \left|\frac{\Delta m_{\rm fly} - \Delta m_{\rm fly}^{\rm max}}{\Delta m_{\rm fly}}\right| \approx 8\frac{a_2 e_2}{\Delta a} \approx 10.6\frac{M_{\bullet,6}}{m_2^{1/3}}\left(\frac{e_2}{0.03}\right)\left(\frac{\Delta a}{R_2}\right)^{-1},
    \label{eq:fecc}
\end{equation}
where in the final approximate equality we have Taylor expanded in the limit $e_2 \ll 1$ and taken $a_2 = r_{\rm RL}$.  Thus, we see that $f_{\rm ecc} \approx 1$ when $e_2 \approx 0.003$, and $f_{\rm ecc} \approx 10$ when $e_2 \approx 0.03$.  Both of these values are consistent with the maximum residual eccentricity of a stellar EMRI at the beginning of RLOF (Eq. \ref{eq:emax}).  We may therefore expect substantial variation in the peak luminosities of many QPEs, though it is possible that some systems will have experienced greater circularization due to tidal evolution.

We conclude by performing several consistency checks on the mass-loss formalism above:
\begin{itemize}
\item 
From Eqs.~(\ref{eq:epsilon}, \ref{eq:scaleheight}) we have that
\be
\frac{\Delta r}{H} \simeq \frac{\epsilon}{H/R_2} \approx 19\frac{m_1}{r_2}\left(\frac{T_{\rm eff}}{10^{4}\, \rm K}\right)^{-1} \left(\frac{\Delta a}{5R_2}\right)^{-2}.
\ee
Thus, for characteristic separations $\Delta a \sim 5 R_2$ (over which most of the mass-loss from $M_2$ will occur; Eq.~\ref{eq:deltaa_dest}), we see that $\Delta r \gg H$ for $m_1 \gtrsim 10^{-2}$, consistent with the assumption made in using Eq.~(\ref{eq:mdotclose}) for the mass-loss rate.  

\item Eq.~(\ref{eq:mdotclose}) assumes the mass-loss occurs as part of a steady-state outflow (e.g., \citealt{Lubow&Shu76}).  We must therefore check that the timescale over which $r_{\rm H}$ is reduced, $\tau_{\rm fly}$, is long compared to the timescale for mass flow through the nozzle $\tau_{\rm flow} \sim \Delta r/v$, where $v \approx c_s \simeq (kT_{\rm eff}/\mu m_p)^{1/2}$ is the outflow rate near the sonic point.  We find,
\be
\frac{\tau_{\rm fly}}{\tau_{\rm flow}} \sim 2\frac{m_2^{1/3}}{M_{\bullet,6}^{1/3}r_{2}^{2}}\left(\frac{T_{\rm QPE}}{10\,{\rm hr}}\right)\left(\frac{\Delta r}{10H}\right)^{-1}\left(\frac{T_{\rm eff}}{10^{4}{\rm K}}\right)^{-3/2}\left(\frac{\Delta a}{5 R_2}\right),
\ee  
consistent with $\tau_{\rm fly} \gtrsim \tau_{\rm flow}$ for characteristic parameters.

\item
The mass-loss $\Delta m_{\rm fly}$ from $M_2$ during each flyby is likely to be sufficiently rapid for the response of the star to be adiabatic.  Its effect on the structure of $M_2$ is then to increase the radius of $M_2$ by a fractional amount $\Delta R_{\rm ad}/R_2 \approx (1/3)(\Delta m_{\rm fly}/M_2)$ for an assumed adiabatic index $\gamma \simeq 5/3$ (e.g., \citealt{Linial&Sari17}).  Thus, in addition to the bursty mass-loss that occurs during each flyby, an enhanced ``steady'' rate of mass-loss from $M_2$ will occur throughout its entire orbit, once it begins to regularly undergo strong interactions with $M_1$.  

Following Eq.~(\ref{eq:mdotclose}), the ratio of the (enhanced) steady mass-loss rate to that experienced during the flyby due to the gravitational influence of $M_1$, can be estimated as:
\be
\frac{\dot{m}_{\rm steady}}{\dot{m}_{\rm fly}} \sim \left[\frac{\Delta R_{\rm ad}/R_2}{\epsilon}\right]^{n+3/2} \sim \left[\frac{\Delta m_{\rm fly}}{M_1}\left(\frac{\Delta a}{R_2}\right)^{-2}\right]^{n+3/2}
\ee
Thus, insofar as $\Delta m_{\rm fly} \ll 0.04M_1(\Delta a/5 R_2)^{-2}$ (as is satisfied over epochs in which $M_2$ loses most of its mass; Eq.~\ref{eq:deltamdest}), we see that $\dot{m}_{\rm fly} \gg \dot{m}_{\rm steady}$.  The accretion rate onto the SMBH will thus indeed be dominated by the punctuated episodes of mass-loss that occur during the flybys, consistent with the observed large amplitude variability of QPEs.

\end{itemize}

\section{Ablation Mass Loss from Stellar EMRIs in Counter-orbiting Gaseous Disk}
\label{sec:ablation}

MS17 estimated the mass-loss rate of $M_1$ due to ablation from the gaseous SMBH accretion flow (albeit in the slightly different context of gaseous disks from tidal disruption events), finding a minimum destruction time which we can express as (MS17; their Eqs.~40-42):
\be
t_{\rm abl} \sim \frac{R_{1}}{v_{\rm c}}\left[\frac{v_{\rm c}}{v_{\rm esc}}\right]^{-3/2}\left(\frac{\rho_{1}}{\rho_{\rm d}}\right)^{5/4},
\label{eq:tabl1}
\ee
where $\rho_1 = 3M_{1}/(4\pi R_1^{3})$ is the mean density of $M_1$, $v_{\rm c}$ is the relative velocity between the stellar orbit and gaseous disk ($v_{\rm c} = 2v_{\rm K}$ in the counter-orbiting case, where $v_{\rm K} = (GM_{\bullet}/r)^{1/2}$ is the Keplerian velocity), $v_{\rm esc} \equiv (GM_{1}/R_{1})^{1/2}$ is the surface escape speed of $R_1$, and 
\begin{eqnarray}
\rho_{\rm d} \approx \frac{\dot{M}_{\rm d}}{6\pi \alpha r^{2}v_{\rm K}(h/r)^{3}}  \approx 3\times 10^{-12}{\rm g\,cm^{-3}}\, \frac{1}{\alpha_{0.1} M_{\bullet,6}}\frac{m_1}{r_1^{3/2}}\left(\frac{L_{\rm X}}{10^{43}{\rm erg\,s^{-1}}}\right)\left(\frac{h/r}{0.3}\right)^{-3}
\end{eqnarray}
is the midplane density of the gaseous disk of steady-state accretion rate $\dot{M}_{\rm d}$ at the orbital radius $r = a_1$ of $M_1$, where in the second line we have taken $L_{\rm X} \simeq 0.1 \dot{M}_{\rm d} c^{2}$.  Combining results, we can now write Eq.~(\ref{eq:tabl1}) as,
\begin{eqnarray}
t_{\rm abl} \approx 9\times 10^{5}\,{\rm yr}\,\alpha_{0.1}^{5/4}M_{\bullet,6}^{5/12}\frac{m_{1}^{23/24}}{r_{1}^{3/8}} \left(\frac{L_{\rm X}}{10^{42}{\rm erg\,s^{-1}}}\right)^{-5/4}\left(\frac{h/r}{0.3}\right)^{15/4}.
\label{eq:tabl}
\end{eqnarray}
For typical parameters (e.g., $h/r \gtrsim 0.1$, $L_{\rm X} \sim 10^{42}$ erg s$^{-1}$) we have $t_{\rm abl} \gtrsim 10^{3}-10^{5}$ yr, longer than the destruction time of the stars due their own self-interaction (Eq.~\ref{eq:tdest}).


\end{document}